\documentclass[sigconf, natbib=false, balance=false]{acmart}
\usepackage[style=ACM-Reference-Format,backend=bibtex]{biblatex}
\usepackage[utf8]{inputenc}
\usepackage{xcolor}
\usepackage{soul}
\usepackage{multirow}
\usepackage[acronym]{glossaries}
\usepackage{tabularx}
\newacronym{bm}{BM}{biological motion processing}
\newacronym{pld}{PLD}{point-light display}
\newacronym{plds}{PLDs}{point-light displays}
\newacronym{pca}{PCA}{principal component analysis}
\newacronym{svm}{SVM}{support vector machine}

\definecolor{teal}{rgb}{0,0,0}

\usepackage{popets}
	\setcopyright{popets}
	\acmVolume{YYYY}
	\acmNumber{X}
	\acmDOI{XXXXXXX.XXXXXXX}
	\acmISBN{}
	\acmConference{Proceedings on Privacy Enhancing Technologies}
\copyrightyear{2023}
\acmYear{2023}

\begin{document}
   \title{Understanding Person Identification Through Gait}

  \author{Simon Hanisch\textsuperscript{1} \qquad Evelyn Muschter\textsuperscript{2} \qquad Admantini Hatzipanayioti\textsuperscript{2} \qquad \qquad \qquad \qquad Shu-Chen Li\textsuperscript{2} \qquad Thorsten Strufe\textsuperscript{1,3}}
  \affiliation{%
   \institution{\textsuperscript{1}Chair of Privacy and Security, Centre for Tactile Internet with Human-in-the-Loop (CeTI), TU Dresden\\
   \textsuperscript{2}Chair of Lifespan Developmental Neuroscience, CeTI, TU Dresden\\
   \textsuperscript{3}Karlsruhe Institute of Technology (KIT), Germany}
   \city{\textsuperscript{1,2}\{firstname\}.\{lastname\}@tu-dresden.de}
   \country{\textsuperscript{3}thorsten.strufe@kit.edu}
  }

\begin{abstract}
{Gait recognition is the process of identifying humans from their bipedal locomotion such as walking or running. As such, gait data is privacy sensitive information and should be anonymized where possible. With the rise of higher quality gait recording techniques, such as depth cameras or motion capture suits, an increasing amount of detailed gait data is captured and processed. \textcolor{teal}{The i}ntroduction and rise of the Metaverse \textcolor{teal}{ is an example of a potentially} popular application scenario in which the gait of users is transferred onto digital avatars.} As a first step towards developing effective anonymization techniques for high-quality gait data, we study different aspects of movement data to quantify their contribution to gait recognition. We first extract categories of features from the literature on human gait perception and then design  experiments for each category to assess how much the information they contain contributes to recognition success. \textcolor{teal}{We evaluated the utility of gait perturbation by means of naturalness ratings in a user study.} Our results show that gait anonymization will be challenging, as the data is highly redundant and inter-dependent.
\end{abstract}


\keywords{biological motion, privacy, gait, identification, recognition, anonymization}

\maketitle
\renewcommand{\shortauthors}{Hanisch et al.}

\section{Introduction}
Human gait is a biometric trait that can be used to identify persons, infer private information such as sex~\cite{shiqi_yu_study_2009} or age~\cite{Zhou2020}, and is used in the diagnosis of medical conditions.
When compared to person identification via faces, gait has advantages as it can be done from distances at which the face is not yet recognizable or occluded by objects such as face masks. It is believed that distinguishing individuals from afar was an important human survival mechanism in the past\textcolor{teal}{,} as it allowed to recognize if an individual was a friend or foe before the person was close enough to be a \textcolor{teal}{potential} threat~\cite{YovelOToole2016}.

In today's world, we no longer need to rely as much on gait recognition for our own safety. 
However, the ease of gait recognition\textcolor{teal}{,} given near ubiquitous means of capturing and recording people\textcolor{teal}{,} creates novel threats to privacy. Examples from 
China\footnote{https://tinyurl.com/5ya4cwdd [apnews], accessed: 17.08.2022}
show that gait is very much suited to be used for surveillance purposes alongside face recognition. 
As for the future, human gait will increasingly be captured and processed  (cf. Fig. \ref{fig:examples} for recent examples), for example, to realize visions like the Metaverse~\cite{10.1145/2480741.2480751}. 
The Metaverse is an immersive virtual world for which human behavior, including human gait, is captured and then transferred onto digital avatars.
Another example is safe human-robot collaboration~\cite{morato2014toward} in which humans have to be closely tracked \textcolor{teal}{in order} to avoid collisions with robots.
The technology used for these scenarios captures the human motion either via vision-based approaches using cameras 
(e.g., Kinect\footnote{https://azure.microsoft.com/en-us/services/kinect-dk/}) or wearable systems based on inertial measurement units (IMUs) (e.g.,  SmartSuit Pro\footnote{https://www.rokoko.com/products/smartsuit-pro}). 

\begin{minipage}[htb]{0.95\columnwidth}
\begin{figure}[H]
	\includegraphics[width=0.3\textwidth]{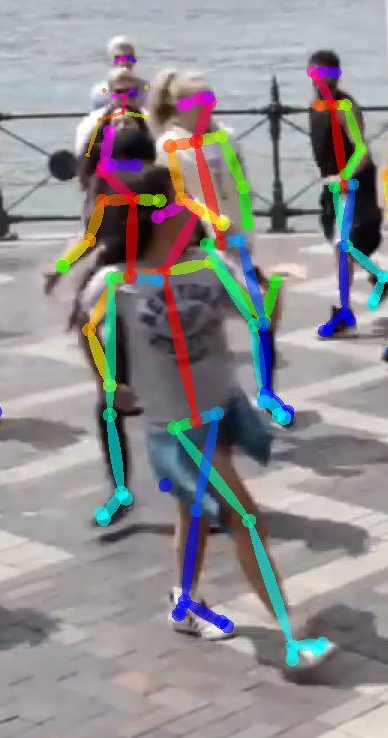}
	\hspace*{0.02\textwidth}	\includegraphics[width=0.3\textwidth]{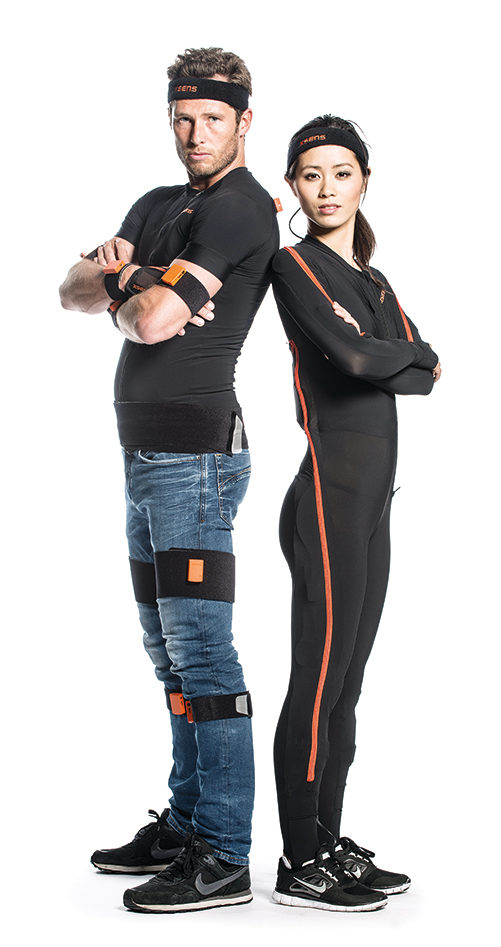}
	\hspace*{0.02\textwidth}
	\includegraphics[width=0.3\textwidth]{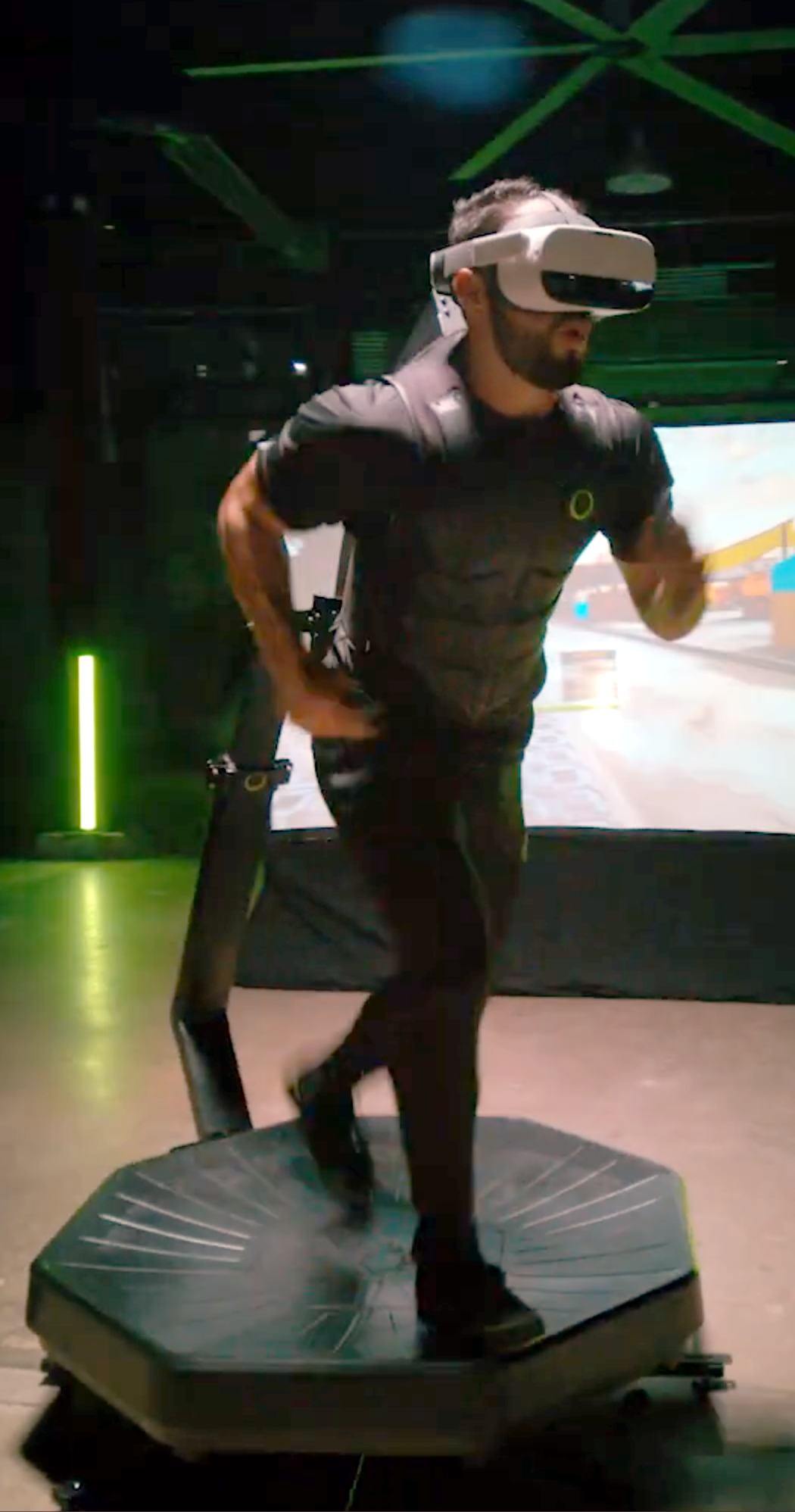}
	\caption{Variety of recent motion capturing devices recording gait of individuals in public and private spaces from video or inertial measurement units (cf. OpenPose, VFXVoice, Virtuix).}
	\label{fig:examples}
\end{figure}
\end{minipage}
\vspace{+0.5cm}
\par
In order to preserve the privacy of individuals, anonymization methods are required to remove or hide sensitive information of their gait. 
We work towards this goal by investigating which features of human gait make individuals identifiable or allow the inference of personal attributes like sex. 
As a starting point for finding categories of features for our computational experiments, we look at the literature on human gait perception\textcolor{teal}{. This line of research has long been a topic of cognitive science and investigates how humans identify other humans by their gait}
(e.g., \cite{ Johansson1973,Troje2005,connor2018biometric}). \textcolor{teal}{We would like to emphasize that we are not interested in designing a novel attack, and we are less interested in investigating the robustness of specific anonymization schemes that are hard to interpret. Instead, we are interested in the question of which features in precise gait data yield identity or attribute disclosure of the individuals, if they coincide with those known from psychological research on gait and person perception---and to which extent they are inter-dependent and hence cannot independently be suppressed/perturbed for anonymization.} 

\textcolor{black}{For a systematic analysis we use machine learning (ML)} 
to get an estimate of how much identifying information exists in gait data. We then design perturbations for each of the feature categories that remove specific features in the gait data\textcolor{teal}{,} to then measure how much the recognition performance drops. 
Where possible, we try to manipulate a feature alone, so that we can estimate how much identifying information the feature contains, as well as how much it shares with other features. \textcolor{teal}{To establish how much utility is retained after anonymization, we performed a user study in which the participants rated the naturalness of the resulting gaits.}
Key contributions of our work are as follows:

\begin{itemize}
    \item human gait feature categorization extracted from the literature of cognitive science;
    \item systematic study of feature contribution for gait recognition for both identification and sex classification;
    \item simple gait perturbation techniques;
    \item \textcolor{teal}{utility evaluation via perceived naturalness of the anonymized gait.}
\end{itemize}
\textcolor{black}{
In the following, we start by first explaining the system model we follow and then review related work from the fields of computer science and cognitive science in Section~\ref{sec:related_work}. This includes the relevant literature in human gait recognition and person perception in the field of cognitive science. In Section~\ref{sec:methodology} we describe our general methodology and the design and implementation of our computational experiments. We present the \textcolor{teal}{ experimental} results in Section~\ref{sec:evaluation}. Finally, Sections~\ref{sec:discussion} \textcolor{teal}{and }\ref{sec:conclusion}  offer a \textcolor{teal} {discussion} and conclusion.
}






\section{Background and Related Work}
\label{sec:related_work}
\textcolor{teal}{In the following section w}e provide background on current motion capturing including potential threats, the human and automatic recognition tasks based on gait, and the state of the art with regards to anonymization and explainability-based analyses of identifying features in gait.

\subsection{System Model and Threats}\label{sec:system_model}
Our main interest is to identify those features in gait data that carry the most information for the identification of individuals, or to infer their attributes, like sex. Our interest is based on the observation that motion is increasingly captured and processed for various purposes.
Consider a user who is using a motion capture system to transfer \textcolor{teal}{their} motions onto a digital representation (e.g., an avatar or digital twin).
Recent examples include various console games, but also the current vision of the \textcolor{teal}{M}etaverse.
These examples require the digital motion to mimic the motions of the user in a natural looking way, resembling \textcolor{teal}{them} as closely as possible.

\textcolor{teal}{The movements of the user are captured locally in real-time}, based on either visual recording or wearables with embedded sensors.
It is then locally pre\textcolor{teal}{-}processed, for example to extract motions from the video or sensory data, and then transmitted to the service provider, who analyses and distributes it to their customers and other users.

Recognizing or identifying a person in such services clearly has social relevance \cite{Cutting1977, Papeo2017} and clinical power for certain disorder diagnostics (e.g., stroke, Parkinson\textcolor{teal}{'s} disease) \cite{Rochester_macrogait,Zhou2020}. 
This immediately implies that there are privacy issues to be considered, both in terms of recognition, for instance for surveillance, but also in terms of attribute inference. 

The {\em service provider} on the one hand will likely be in the position of identifying and linking subsequent observations of an individual based on their account.
However, the processed motion data may yield inferences of attributes that are private and not relevant \textcolor{teal}{to} the service, and the possibility of linking several independent accounts of the same individual.
\textcolor{teal}{On the other hand, }{\em third parties} \textcolor{teal}{e.g.,} other users or platform and advertising customers of the provider, will be able to train recognition models of users in order to re-identify (for instance in auxiliary video data), or also learn sensitive attributes that are not shared voluntarily by the individual.

Hence, we aim to understand which features of the motion provide high predictive power for identity or attribute disclosure.
This would facilitate \textcolor{teal}{the development of} local privacy-enhancing technologies for the pre-processing of data \textcolor{teal}{prior to} their transfer.
Understanding, which features carry information required for the motion to be perceived as natural will help strike useful privacy-utility tradeoffs.


\subsection{Motion Analysis}
\label{sec:mocap_anal}
Humans can recognize and identify biological traits visually through the use of static information such as shape or other cues. Biological motion is one additional important factor. Human newborns, infant monkeys, and even freshly hatched, visually naïve chicks show a preference for motion cues that move in motion patterns suggesting animacy \cite{lorenzi2017dynamic}.

\begin{figure}[h]
\centering
      \includegraphics[width=0.15\textwidth]{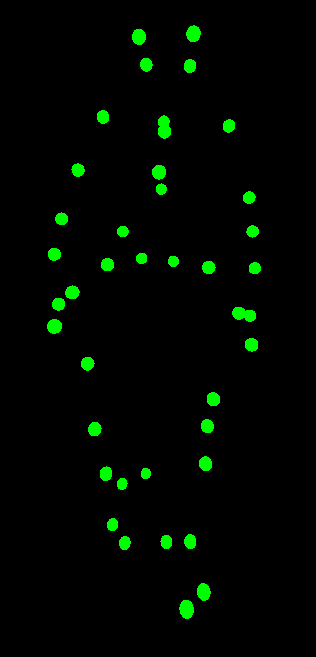}
\caption{Example of a human walker represented as a point-light display (PLD).}\label{img:point-light-display}
\end{figure}

An established and reliable method of investigating biological motion processing without other potential sensory information or cues (e.g., color, texture, or form-based features such as facial configurations, hairstyle, or clothing) is the use of \acrfull{plds}~\cite{Johansson1973} (see Fig.~\ref{img:point-light-display}). 
This method of using impoverished moving dot displays helps to isolate motion information from other cues. 
Thereby, a small number of dots represent the head and major joints of a human body in various scenarios such as social interactions~\cite{Bellot2021, Manera2010} or\textemdash as the focus of this work\textemdash during gait~\cite{Troje2005}.
Indeed, many identifying features of a human walker can be recognized from such \acrshort{plds}.

\begin{figure}[!h]
\centering
      \includegraphics[width=0.4\textwidth]{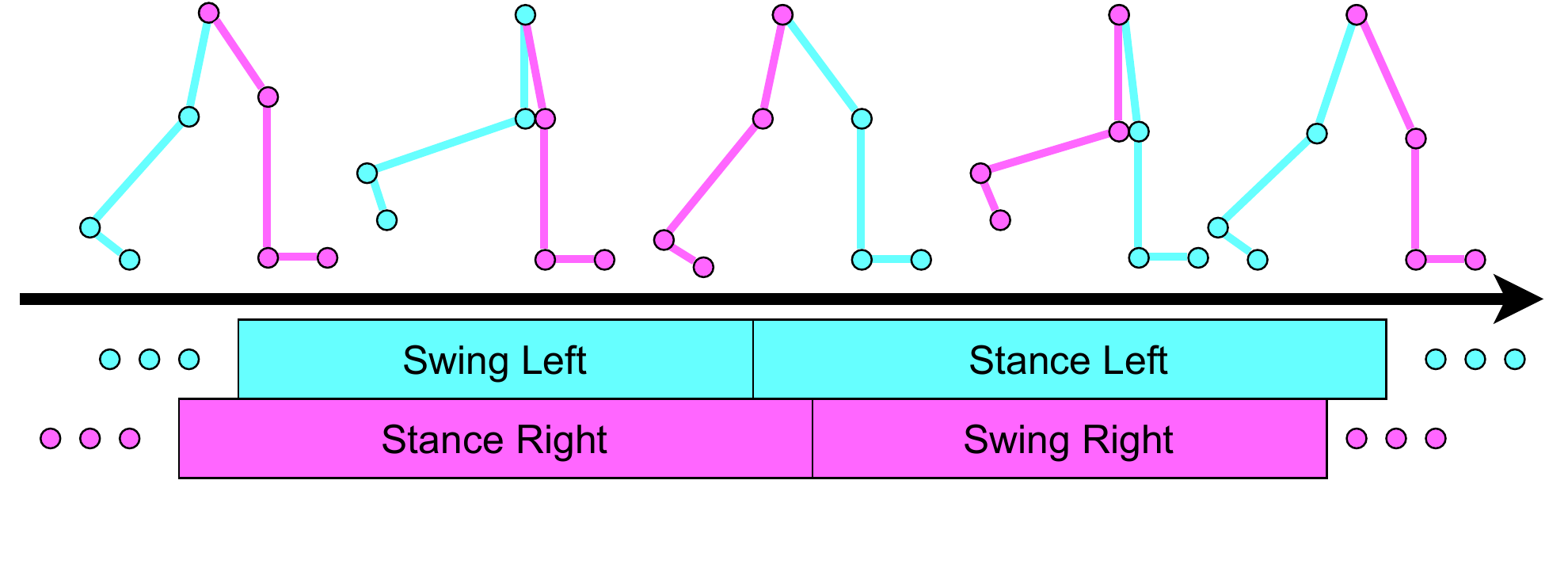}
\caption{A schematic representation of the gait cycle.}\label{img:gait_cycle}
\end{figure}

\textit{Gait analysis} is the study of human locomotion (walking and running) and defines walking as a series of gait cycles. A \textit{gait cycle (stride)} is the period when one foot contacts the ground to when that same foot contacts the ground again (see Fig.~\ref{img:gait_cycle}). Each gait cycle has two phases: the stance phase, when the foot is in contact with the ground; and the swing phase when the foot is not in contact with the ground. \textcolor{black}{ In vision-based gait analysis such as PLDs\textcolor{teal}{,} kinematic data such as position and velocity are captured. These are used to relate motion parameters such as joint angles, joint velocity\textcolor{teal}{,} and center of mass with qualitative gait parameters— the general prototypical bipedal motion characteristics such as step length, walking speed, pace and rhythm of steps, stance\textcolor{teal}{,} and swing times \cite{Rochester_macrogait} as well as arm swing, vertical head movement, pelvic rotation, and the extension and flexion of limbs and shoulders. As walking consists of a series of multiple gait cycles, gait data typically also contains fluctuations. These are small variations such as asymmetry and variability in step and stance time, step velocity, or step length \cite{Rochester_macrogait}. These gait parameter\textcolor{teal}{s} can subsequently be used to extract statistical features (e.g., mean, standard deviation, skewness) for gait analysis.}

\subsection{Human Gait Perception} \label{sec:humanperception}

Human observers have no trouble making sense of the very limited information presented through \acrshort{plds}’s disconnected dots, representing actions such as the specific categorical biological motion content \cite{YovelOToole2016,Papeo2017} of human gait (walking or running).
\textcolor{black}{Research has suggested that humans are especially tuned to recognizing conspecifics and this preference is likely to already emerge in the visual system.} Psychophysical and neuroscientific studies have shown that at least two processes play a role in person perception (here defined as the recognition of human bodies and their biometric features based on vision). 

On the one hand, form-from-motion \textcolor{black}{cues}~\cite{Simhi2020VR,YovelOToole2016} are cues that are rooted in basic perceptual abilities to see structure from motion. That is, the shape and form of an object or person are revealed more clearly through motion. The human visual system benefits from the motion direction information in order to extrapolate the overall shape of an object or person.
These cues provide time-invariant information about body form by enhancing the shape presentation of a person \cite{YovelOToole2016} and are susceptible to violations of the hierarchical body form structure \cite{Talia2016,Peng2020} as well as to inversion effects \cite{Gandolfo2020}. \textcolor{black}{That is, inverting a body in the image plane (i.e., placing it upside-down) results in perceptual impairments. Previous research has proposed that inversion is deleterious to normal human whole-body perception, causing observers instead to rely on local part-based visual features \cite{Gandolfo2020}.} 
Evidence for first order configural processing has shown that visual perception of bodies is mediated by spatial configurations of body parts, such as the general body layout (e.g., legs attached to the hip, arms attached to the shoulders), and thus providing intact spatial configurations of bodies \cite{Talia2016}.

On the other hand, dynamic identity signature~\cite{Simhi2020VR, YovelOToole2016}
, describes the idiosyncratic motion pattern of an individual. These features describe the change over time during a walking cycle and rely on nuanced, person-specific motion variations (e.g., the way \textcolor{black}{Charlie Chaplin} walks). 
Furthermore, research has provided evidence for a two-stream processing of biological motion perception in the brain. 
That is, biological motion perception relies on\textemdash both dynamic and static features through\textemdash motion processing in the dorsal pathway (i.e., area V5 of visual cortex in the brain) in combination with bodily form and appearance information in the ventral pathway in the brain (see Peng et al.~\cite{Peng2020} for further details on visual recognition of biological motion). \textcolor{black}{In addition to action recognition, human observers are able to identify soft biometric features of actors in \acrshort{plds}, including sex \cite{Montepare1988_age,Gandolfo2020,Kozlowski1977_gender}, age \cite{Zhou2020,Montepare1988_age}, weight \cite{morrison2018something}, height \cite{morrison2018something}, handedness \cite{loffing2016tennishand}, in addition to attractiveness \cite{Montepare1988_age,morrison2018something}, identity \cite{Cutting1977}, emotions \cite{Montepare1988_age,Johnson2011} and causal intentions \cite{Manera2010}. Specifically, Kozlowski and Cutting \cite{Kozlowski1977_gender} showed that the biomechanical factor center of moment, which is derived from the relative movement of both -- the shoulders and hips, plays a crucial role in sex perception in \acrshort{plds}. }
 
\textcolor{black}{Finally, person recognition depends on familiarity and might take time for the human observer to learn, but is useful for recognizing a familiar person from a distance \mbox{\cite{Cutting1977, Troje2005}}.} \textcolor{black}
{Studies have shown that human observers are more sensitive to \mbox{\acrshort{plds}} of themselves and friends \mbox{\cite{Cutting1977,Loula2005}}, or could learn to identify a small number of individuals based on their motion \mbox{\cite{Troje2005}}.}
Guided by these insights, we \textcolor{black}{aim} to investigate if the removal of the \textcolor{black}{certain} features will reduce recognition rates, and to which extent.
\textcolor{black}{Specifically, we focus on the features that are easy to extract from existing data sets. Namely, macro and micro features (i.e., statistical features, see Sect. \ref{sec:mocap_anal}), perturbations of intact bodies in natural spatial configurations, as well as dynamic (i.e., temporal information) and static (i.e., structural) features. Section \ref{sec:obfuscate_features} will describe the specific features used in the present study in detail.}

\subsection{Automatic Gait Analysis}
Current human movement analyses are based on biometric measurements and motions. They are captured vision-based, or using wearables with integrated inertial measurement units (IMUs).
A gait cycle is thereby composed of a chain of individual 2D (video) or 3D (optical marker/IMU tracking) samples at each given time point (pose).

Gait recognition using machine learning models is most commonly based on video data~\cite{wan_survey_2019}.
Video, providing rich information about subjects, facilitates high recognition rates and hence is frequently used for surveillance purposes~\cite{balazia_you_2017}. 

Also explicit motion capturing frequently uses video: High-quality vision-based motion capturing uses specialized cameras to track reflective markers on the subject's body. 
The position of these markers is later reconstructed into 3D position time series, converted into joint angles as a function of time, and subsequently analyzed according to specific research or clinical needs. 
However, recently, approaches of using a single commodity camera in combination with keypoint detection algorithms and neural networks (e.g., Open Pose or DeepLabCut, cf. Fig. \ref{fig:examples}) have generated convincing results \cite{kidzinski2020deep, mehdizadeh2021concurrent,mathis2018deeplabcut}.

Gait recognition is also possible based on motion capturing data~\cite{balazia_gait_2018}. 
Indeed, even simple kinematic features obtained from IMU systems (e.g., position, velocity, and acceleration-based features) or kinetic data from force plates and electromyography (e.g., ground or muscle force parameter) have been shown to yield high recognition rates (see~\textcite{connor2018biometric} for an overview). 

Anonymizing individuals in video surveillance footage for multiple moving object detection and tracking algorithm (e.g., human action tracking) by representing their bodies as simplified objects such as \acrshort{plds} thus cannot protect their identities.
Further, gait can also be used to infer personal attributes like sex \cite{shiqi_yu_study_2009} and age~\cite{Zhou2020, eskofier_marker-based_2013}. Being interested in those gait features that carry information for identification and attribute disclosure of individuals, \textcolor{teal}{in the present work we rely on marker-based motion capture data as it is considered the gold standard in the field.}


\subsection{Existing Anonymization Approaches}

As protection against gait recognition, gait anonymization recently became an active field of research.
\textcolor{teal}{However, in the present paper w}e are not interested in evaluating their efficacy, \textcolor{teal}{but} rather want to establish \textcolor{teal}{an} initial understanding as to which part of gait data carries identifying information.
Yet, we use the state of the art on anonymization for guidance.

For example, \textcolor{teal}{Ivasic-Kos et al.}~\cite{ivasic-kos_person_2014} anonymized activity videos by blurring the data. 
This clearly \textcolor{teal}{retains} the gait information in the data and does not facilitate insights into its influence on identification.
More sophisticated methods use neural network approaches to perturb the video recordings of gait and then generate a new natural-looking gait sequence~\cite{tieu_approach_2017, hirose_anonymization_2019}.
While manipulating, and hence implicitly providing indicators about the influence of different properties in gait, neither approach investigates the actual influence but rather provides ad-hoc anonymization.
For accelerometer-based data, it was proposed to add noise directly to signals collected \textcolor{teal}{via smartphones}~\cite{matovu_jekyll_2018}.
This perturbation may have some effect given singular \textcolor{teal}{smartphone-based} measurements.
However, it can be removed from the data, as Wang et al.~\cite{wang21why} have shown for trajectories, and it does not provide insights as to which details of the data are identifying the users.

\subsection{Investigations using Explainability}
The field of explainable machine learning~\cite{samek19explainable, arya19one} focuses on the interpretability of learned machine learning models.
It helps to determine those features that are important for the classification, for instance of gait recognition systems.

A common approach is layer-wise relevance propagation (LRP) \cite{bach_pixel-wise_2015}.
It is applied to neural networks (NN) to find the most relevant features in the input data by tracing the classification back through the layers of the NN. 
Another approach is to perturb areas in the input data in order to find the area that results in the largest classification performance reduction~\cite{fong_interpretable_2017}. 
The corresponding insights are limited to implicit indicators to regions of the input data that influence the classification accuracy. 
They do not provide insight as to the explicit semantic feature of the data, in our case\textcolor{teal}{—}the different characteristic features of gait, that contribute to identification of the individuals.

Approaches for explainable machine learning have also been used to analyze gait patterns for clinical analyses~\cite{slijepcevic_explanation_2020}. Horst et al.~\cite{horst2019explaining} used LRP to study which part of the gait cycle is relevant to a non-linear machine learning model to recognize an individual.
Furthermore, \textcite{connor2018biometric} investigated which features in gait data make people identifiable. These approaches test trained models for effects of \textcolor{teal}{the} removal of features, without retraining them. The question remains, whether other features were considered unnecessary and \textcolor{teal}{thus} ignored by the initially trained model, \textcolor{teal}{as all of their information is also contained in the features used by the model.}

\textcite{alvarez_melis_towards_2018} proposed faithfulness as an important metric for evaluating explainable machine learning. \textcolor{teal}{Faithfulness is obtained} by removing/perturbing the feature and then measuring the drop in classification performance.
\textcolor{teal}{While their} investigation does not shed light on the contribution of explicit gait features, \textcolor{teal}{it does} provide instructive ideas for our analysis.

It can be concluded that several studies have investigated gait recognition by both humans and machines, as well as implicit approaches to anonymize gait.
\textcolor{teal}{However, so far t}he question \textcolor{teal}{of} which explicit features of gait have predictive power for identification tasks\textcolor{teal}{—}and hence have to be perturbed or removed for anonymization\textcolor{teal}{—}has not been investigated systematically.

\section{Methods}\label{sec:methodology}

\begin{figure}[!h]
\centering
\includegraphics[width=0.45\textwidth]{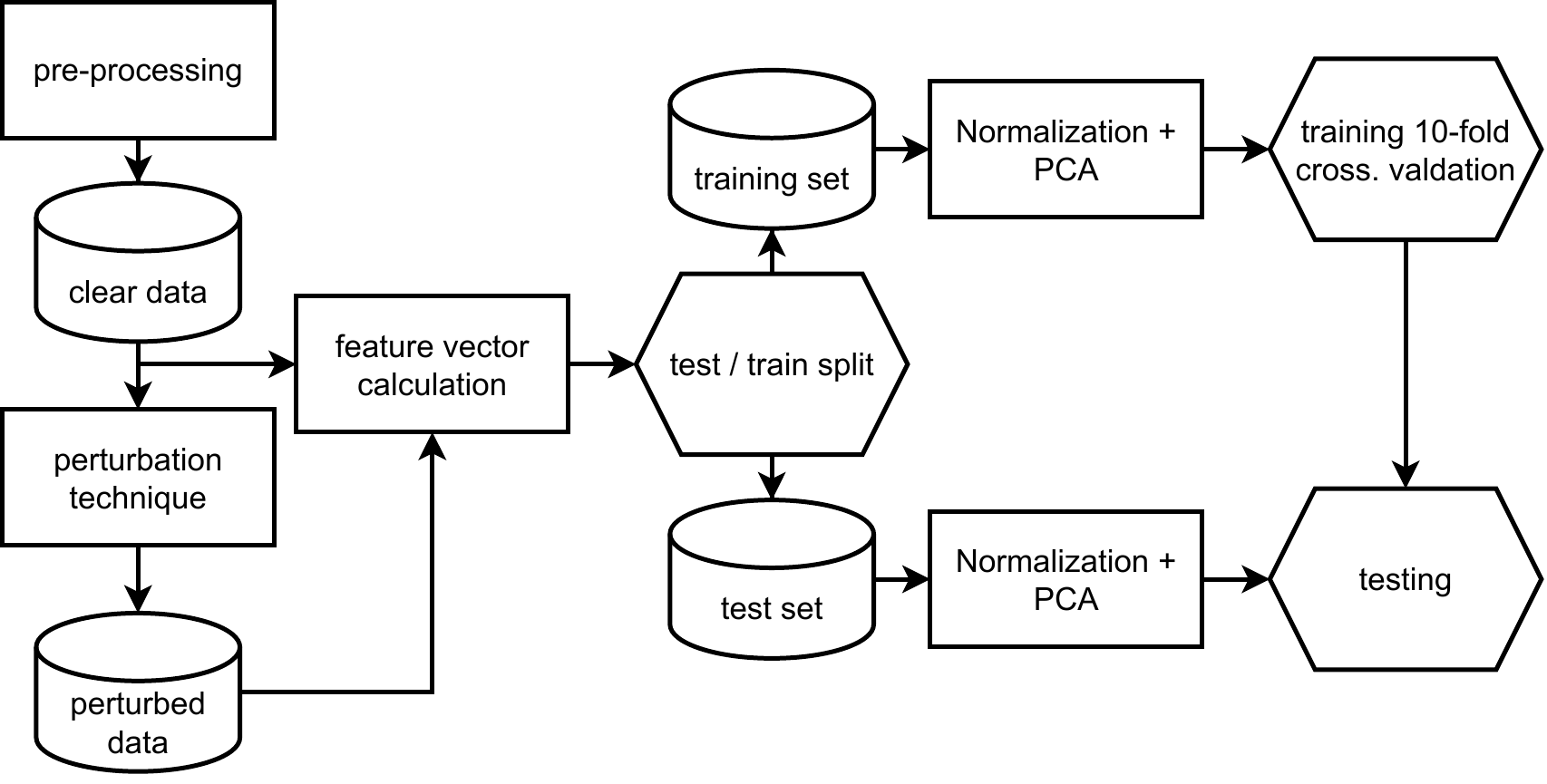}
\caption{The full data processing pipeline.}\label{img:data_processing}
\end{figure}

The question we sought to answer is how much specific features in the data contribute to the overall gait recognition performance of identity and sex using machine learning. 
Our overall approach was to first train \& test a gait recognition system for each of the recognition goals on clear data to obtain baseline accuracy. Next we obfuscated a feature at a time in the data by either perturbing or removing it to investigate its impact on anonymization. 
We then repeated the training \& testing process and report the resulting recognition performance. 
The difference in recognition between baseline and perturbed data gives us the unique \textcolor{teal}{amount} a feature contributes to the overall recognition performance.
Further, we also measured each feature independently from the other features. However, this is only possible for features we can remove from the data and not for features we only perturb. The full process is show in Fig.~\ref{img:data_processing}.


In the following, we provide details about the  data set, applied feature perturbations, the implementation of the recognition system, \textcolor{black}{and utility evaluation}.

\subsection{Data \textcolor{teal}{S}et}
As our main goal was to understand the important features of human gait\textcolor{teal}{,} we chose the highest quality of gait data and used optical 3D marker-based motion capture data for our experiments. This data is considered the gold standard for motion capturing and is recorded using multiple infrared cameras which capture markers on the anatomic landmarks of participants. The benefit of the 3D representation is that there is no dependency on the recording angle like in video recordings. The data consists of multiple samples per participant which are a time series of poses. 
Each pose contains the 3-dimensional coordinates of each marker (placed on the participant) at a given point in time \textcolor{black}{ (i.e., \mbox{\acrshort{plds}}). The data is also more appropriate for our purpose, as we focus on gait features in the absence of potential additional information (e.g. video recordings).}
\setstcolor{black}{
Furthermore, one can generate \mbox{\acrshort{plds}} from the data, using the tracked marker positions.} 

We used the open-source data set by Horst et al.~\cite{horst2018dataset, horst2019explaining} which consists of full-body kinematic and kinetic data of 57 individuals (29 female, 28 male; $23.1\pm2.7$ years; $1.74\pm0.10$m; $67.9\pm11.3$kg). An optical motion capture system and a full body marker set (62 markers corresponding to anatomical landmarks), as well as two force plates, recorded self-paced walking trials at 250Hz (motion capture) and 1000Hz (force plates). \textcolor{black}{For each participant 20 samples containing a full gait cycle have been recorded} (for further details on the data acquisition protocol see  \textcolor{teal}{Horst et al.} \cite{horst2019explaining}).

\subsection{Data \textcolor{teal}{P}re\textcolor{teal}{-}processing}
\textcolor{black}{
Following the methodology of Horst et al.~\cite{horst2019explaining}, we trim\textcolor{teal}{med}
the gait samples to contain only a single stride by using the kinetic force signals of the force plates, using a ground force threshold of 20N. This way all samples are aligned and start at the same point in the gait cycle. The data was then normalized, in order to obtain an equal number of poses for each individual, by resampling each sample to 100 frames.
Each frame represents one discrete pose of the individual while walking, the 100 poses then constitute one stride.}

\subsection{Retained and \textcolor{teal}{M}asked Features}\label{sec:obfuscate_features}

\textcolor{black}{
We based our feature categories on previous work in gait analysis and human perception as described in Sections ~\ref{sec:mocap_anal} and ~\ref{sec:humanperception}. The category name always gives the kind of feature we sought to retain \textcolor{teal}{,} while the perturbation techniques employed are aimed at removing the other features from the data. For each technique, we strove to design an inverse perturbation technique that only removes the specific feature\textcolor{teal}{,} while keeping all the others (micro vs. macro, dynamic vs. static). This way we sought to understand how much each feature contributes to the overall recognition rate and if it contains information that is unique to this feature. Since there is interdependence between features, some of the features are partially overlapping for example the walking frequency is dependent on the walking speed and the length of the legs. Table~\ref{tab:features} gives a brief overview, while the used and obfuscated features are described in detail in the following.\\
Our \textbf{macro} features describe the general characteristics of the walker, such as walking speed, general movement trajectories, walking amplitude, the most significant parts of the walker positions, and overall body parts. Its counterparts are the \textbf{micro} features which contain the small variations of the trajectories that remain when the overall trajectories are removed, the walker without its walking speed and step length equalized over all walkers, the least significant parts of the walker positions, and individual body parts. Besides macro and micro, we also investigated the dynamic parts of the gait motion. For this we have two contrary feature categories static and dynamic. The \textcolor{teal}{\textbf{static}} features contain the time-invariant features, such as the average pose of the walker, or the first pose of the walker. The \textcolor{teal}{\textbf{dynamic}} features contain the features describing the motion of the walker, including the differences between the recorded poses, and walker where the static frame (body proportions) has been removed. The following section describes the used perturbation techniques for each feature category. The parameter values have been chosen to match the used data set. In the end, we briefly detail how we combined the perturbation techniques.
}


\begin{table}[!h]
\caption{used and obfuscated features}
\centering
\resizebox{8.5cm}{!} { 
\begin{tabular}{l|l|l|l|l|}
\cline{2-5}
 &
  \textbf{Macro} &
  \textbf{Micro} &
  \textbf{Static} &
  \textbf{Dynamic} \\ \hline
\multicolumn{1}{|l|}{\textit{definition}} &
  \textit{\begin{tabular}[c]{@{}l@{}}Step length,\\ walking speed,\\ cadence\end{tabular}} &
  \textit{\begin{tabular}[c]{@{}l@{}}asymmetry and\\ variability \\ in the macro features\end{tabular}} &
  \textit{\begin{tabular}[c]{@{}l@{}}Shape and\\ general body layout\end{tabular}} &
  \textit{\begin{tabular}[c]{@{}l@{}}Time course\\  of changes\end{tabular}} \\ \hline
\multicolumn{1}{|l|}{Perturbation 1} &
  Remove variations &
  Remove trajectories &
  Static pose  &
  Motion extraction \\ \hline
\multicolumn{1}{|l|}{Perturbation 2} &
   &
  \begin{tabular}[c]{@{}l@{}}Amplitude/ \\ frequency equalization\end{tabular} &
   Resampling &
   Normalization \\ \hline
\multicolumn{1}{|l|}{Perturbation 3} &
  Coarsening macro &
  Coarsening micro &
   &
   \\ \hline
\multicolumn{1}{|l|}{Perturbation 4} &
  Remove body parts &
  Keep body parts &
   &
   \\ \hline
\end{tabular}
} 
\label{tab:features}
\end{table}
We provide a sample video rendering\footnote{\textcolor{black}{https://github.com/anonPETs2023/PETs2023}}
of all perturbation methods alongside this paper. 

\subsubsection{Macro Features}
The macro features keep the overall characteristics of the walker and remove its smaller variations from the data. We used three perturbation techniques for this: remove variations, coarsening macro, and remove body parts.\\
\textit{Remove variations}: In order to extract the ideal trajectory from the gait data we removed the small variations that deviate from the ideal trajectory. The ideal trajectory is here calculated by two different methods\textcolor{teal}{:} either using a moving window on the marker poses and then calculating a rolling average\textcolor{teal}{,} or an interpolation. The difference between the two is that the rolling average takes all poses in the window to calculate an average, while the interpolation only uses the poses at the edge of the moving window. 
The moving window size is given as the distance to the pose which is calculated and is either one \textcolor{teal}{or three} additional pose\textcolor{teal}{(s)} before and after e.g., spanning three poses in total or spanning seven poses in total, respectively. This strategy follows a similar idea to low-pass filtering, as it retains the main movement but removes detailed deviations.\\
\textit{Coarsening macro}: As we were interested in the most significant information of the walker position\textcolor{teal}{,} we removed the least significant part of each marker position in a pose for all poses. The effect is that the grid on which the walker moves is becoming more coarse. We removed all digits either below the thousandth (1000) or the hundredth digit (100).\\
\textit{Remove body parts:} We measured how much an individual body part (head, torso, hip, arms, legs) contributes to the overall recognition performance. This was done by removing the body part from the data by setting its marker positions to zero.

\subsubsection{Micro Features}
The micro features are the counterparts to the macro features\textcolor{teal}{. H}ere we kept the small variations of the gait cycle and the least significant parts of the marker positions.\\
\textit{Remove trajectories}: \textcolor{teal}{Contrasting} remove variations\textcolor{teal}{,} we removed the ideal marker trajectories from the data by calculating the ideal trajectory as described in remove variation via either rolling average or interpolation with a window size of 1 or 3. The ideal trajectory was then subtracted from the real trajectory\textcolor{teal}{,} which leaves us with the distances of the ideal marker positions to the real ones. 
This strategy resembles high-pass filtering, as it removes the main movement and only retains the minor specifics of the current sample.\\
\textit{Coarsening micro}: We eliminated the most significant part of the walker positions by removing the most significant parts of each marker position value. We removed all digits above the hundredth (100), tenth (10), or first digit (1) position\\
\textit{Keep body part:} We measured how much recognition performance the individual body parts have alone without the rest of the body. All \textcolor{teal}{remaining other }body parts are set to zero.\\
\textit{Amplitude/Frequency equalization}: The walking amplitude and frequency were equalized between all individuals to perturb their influence on the recognition. Informed by previous studies~\cite{Troje2005}, we calculated a gait representation of each individual by using the average pose, the first four components of a principal component analysis (PCA), and a sinus function fit on these components to represent the gait cycle of a person. We then equalized the frequency or amplitude of the fitted sinus function by means of the group-level average.

\subsubsection{Static Features}
The static features capture the time-invariant features of the walker by removing the dynamic part of the gait motion. We therefore kept the proportions of the walker.\\
\textit{Static pose}: We used only an average pose or the first pose of each sample, thus removing the dynamic component of the gait data.\\
\textit{Resampling}: We downsampled the data to 10 frames, and therefore removed most of the dynamic content from the data.

\subsubsection{Dynamic Features}
The dynamic features are the counterpart to the static features and aim to only retain the dynamic part of the motion.\\
\textit{Motion extraction}: Instead of using the individual poses\textcolor{teal}{,} we used their difference (i.e., keeping only the variations between poses) and hence removed the static features.\\
\newpage
\noindent\textit{Normalization}: We normalized the static features in a sequence by either normalizing the height axis (y-axis), all axes or normalizing each dimension over the entire sequence of poses.






\subsubsection{\textcolor{black} {Combinations of Perturbations}}

\textcolor{black}{Besides evaluating each of the features alone\textcolor{teal}{,} we also investigated their combinations. Two perturbation techniques were combined by applying them sequentially to the data. Due to some techniques (first pose, average pose) not returning a time series\textcolor{teal}{,} not all combinations of methods are possible. As the overall number of combinations is quite high\textcolor{teal}{,} we focused on representatives of each class of features. We picked those representatives by their anonymization impact on the data.}


\subsection{Recognition \textcolor{teal}{S}ystem}

To test the impact of omitting features from human gait, and hence their contribution to inference, we implemented a gait recognition system.
It is based on the system by Horst et al.~\cite{horst2019explaining} using Python 3.8.3~\cite{python383}, Scikit-learn 0.23.1 \cite{scikit-learn}, and NumPy 1.18.5 \cite{harris2020array}. We used two feature vectors to represent a data sample: \textit{flatten} which concatenates all poses of a sample into a single vector, and \textit{reduced angles} which first calculates a reduced representation of 17 markers representing the main body parts and then calculated 10 joint angles from this representation.

Next, the data was split into train (75\%) and test (25\%) data. Here we differentiated between the identity and sex recognition. For identity recognition, we split the samples for each identity so that we have every identity in both sets. While for sex recognition we split the samples identity-wise, making sure that every identity is only in one of the sets. \textcolor{black}{ We did so to make sure that the classifier cannot learn the identity to perform sex recognition.} Following the split, we then scaled the data in each set by subtracting the mean and then scaling with the standard deviation before we performed a principal component analysis (PCA) to reduce the dimensions of the samples. As a classifier, we used a support vector machine (SVM) using a radial basis function (RBF) kernel. For the training of the SVM we used 10-fold cross-validation with the train set before we tested the best performing model on the test set. In order to account for the random splitting of the data, we ran the entire process 10 times.

\subsection{\textcolor{black} {Utility} } \label{sec:meth_utility}

\textcolor{black}{
Besides investigating the identity and sex recognition performance of our features we also sought to understand how much the features contribute to the utility of envisioned applications. As our use case (see Sec.~\ref{sec:system_model}) is to transfer the gait motion onto a digital avatar, the goal is to retain as much naturalness in the motion data as possible. In order to measure the corresponding effect, we performed an online survey with 22 human participants (13 male, age: 18--60 years) 
which we asked to rate the naturalness of the perturbed gait sequences. Participants were shown renderings of two gait sequences for each perturbation in which the walkers (one male and one female walker, individually) were shown from the side 45 degrees rotated around the z-axis towards the camera. The renderings are identical to the example videos we provided in Section~\ref{sec:obfuscate_features}. All sequences were shown in random order. The participants then rated on a scale from 1 (worst) to 5 (best) how natural looking the gait sequence appeared to them.} \textcolor{teal}{The survey data collection is under the umbrella of the project ("Privatsphäre von Körperbewegungen") approved on 30.09.2021 by the ethical committee of KIT and was conducted in accordance with the Declaration of Helsinki. The survey data was collected in anonymized form.}

\section{Results}\label{sec:evaluation}

In this section we \textcolor{black}{present the results of} our obfuscation experiments, by reporting the recognition performance of the chosen feature categories. The results for identity and sex recognition in two contrasting feature categories (macro vs. micro, dynamic vs. static) are reported each. Note, that we report the body part removals (body parts vs. rest body) separate from the macro and micro features for easier comparability.



\textcolor{black}{As we conducted recognition experiments and the classified classes (for both identity and sex recognition) have nearly the same number of samples per class, we selected accuracy as our metric. Accuracy is defined as the number of correctly classified samples divided by the number of all classified samples. 
In Section~\ref{sec:methodology} we described the two feature vectors we used in our recognition system\textcolor{teal}{. S}ince we were interested in how much identifiable information remains in the data after the perturbation has been applied\textcolor{teal}{,} we always report the best performing feature vector.} 

\subsection{Macro vs. \textcolor{teal}{M}icro \textcolor{teal}{F}eatures}

\begin{figure}[!h]
\centering
      \includegraphics[width=0.475\textwidth]{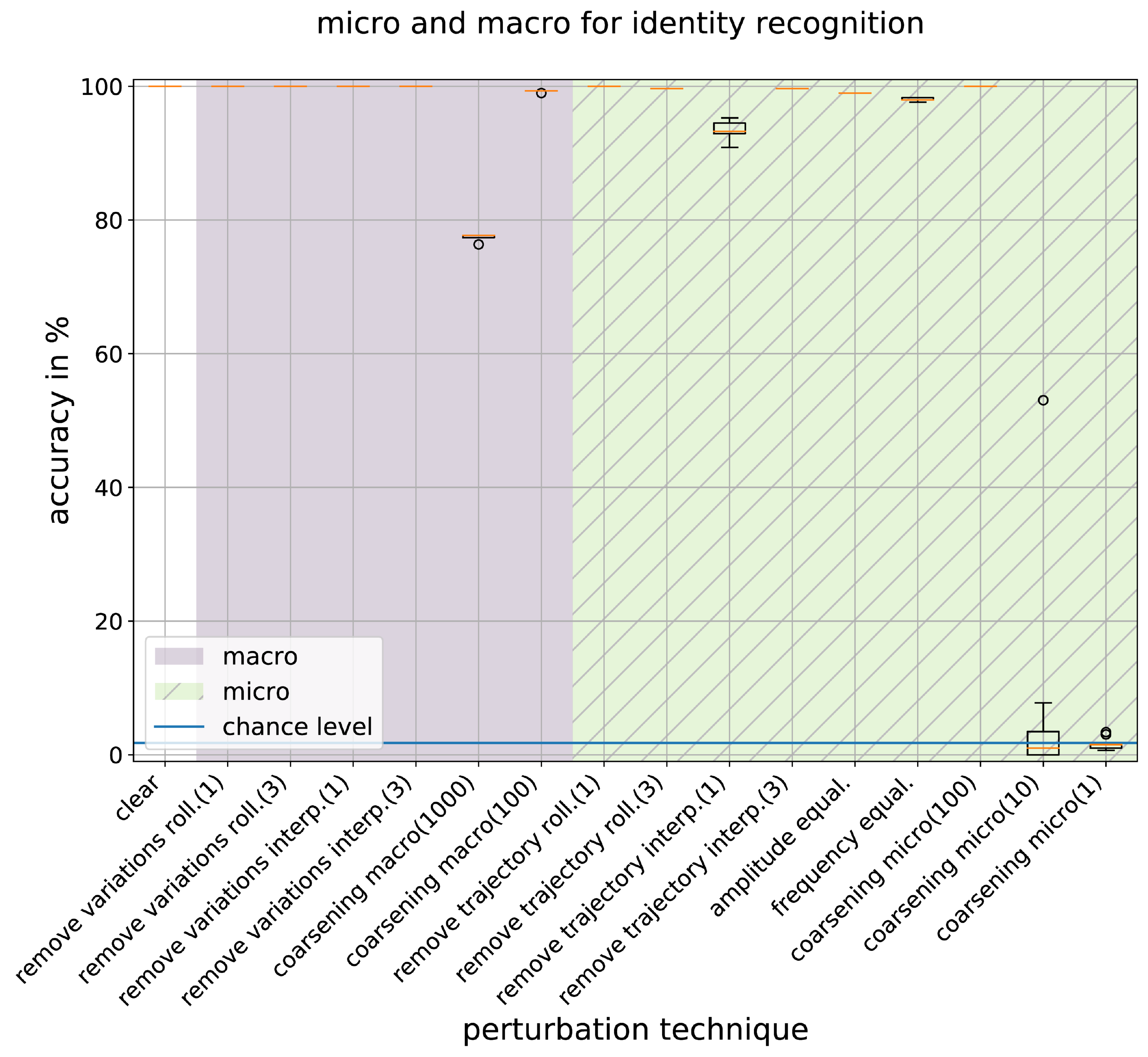}
\caption{Boxplot of accuracy results for micro and macro features for identity recognition given in percent.}
\label{img:micro_vs_macro_identity}
\end{figure}

\begin{figure}[!h]
\centering
      \includegraphics[width=0.475\textwidth]{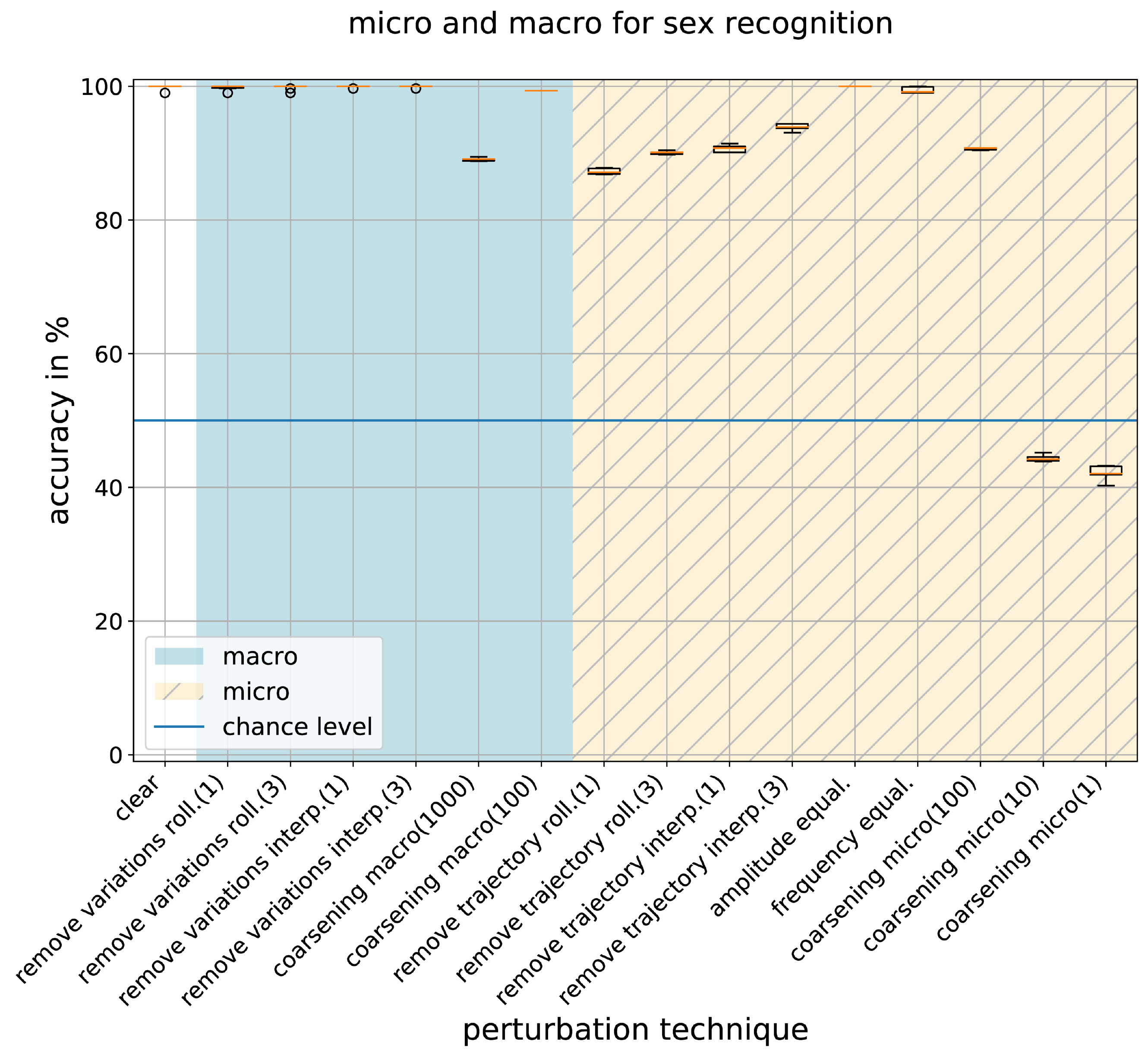}
\caption{Boxplot of accuracy results for micro and macro features for sex recognition given in percent.} 
\label{img:micro_vs_macro_gender}
\end{figure}

\textcolor{black}{
We start by comparing macro to micro features. For both, identity (Fig.~\ref{img:micro_vs_macro_identity}) and sex recognition (Fig.~\ref{img:micro_vs_macro_gender}), we can see similar effects for the macro features: The variation removal via rolling average and interpolation shows no effect on the accuracy. The coarsening of all digits below the 100th digit has no effect\textcolor{teal}{,} while coarsening from the 1000th digit position leads to a drop in accuracy for sex recognition to 91\% and identity recognition to 77\%. For the micro features, we see a difference between identity and sex recognition. Only trajectory removal using an interpolation window of 1 drops the accuracy of the identity recognition\textcolor{teal}{,} while all of the others lead to a drop in sex recognition to about 90\%. For the micro coarsening methods we again see that identity recognition is not affected by coarsening everything higher than the 100th digit\textcolor{teal}{,} while for sex recognition we see a drop of accuracy to 90\%. Then coarsening the digits above the 10th digit leads \textcolor{teal}{for }both\textcolor{teal}{, identity and sex recognition,} to chance level accuracy. The results show that sex recognition is more dependent on the macro feature than on the micro features\textcolor{teal}{,} while the identity can be perfectly inferred from both of them.
}


\subsection{Individual \textcolor{teal}{B}ody \textcolor{teal}{P}arts in \textcolor{teal}{I}solation vs. \textcolor{teal}{R}educed \textcolor{teal}{W}hole \textcolor{teal}{B}odies }
\begin{figure}[!h]
\centering
      \includegraphics[width=0.475\textwidth]{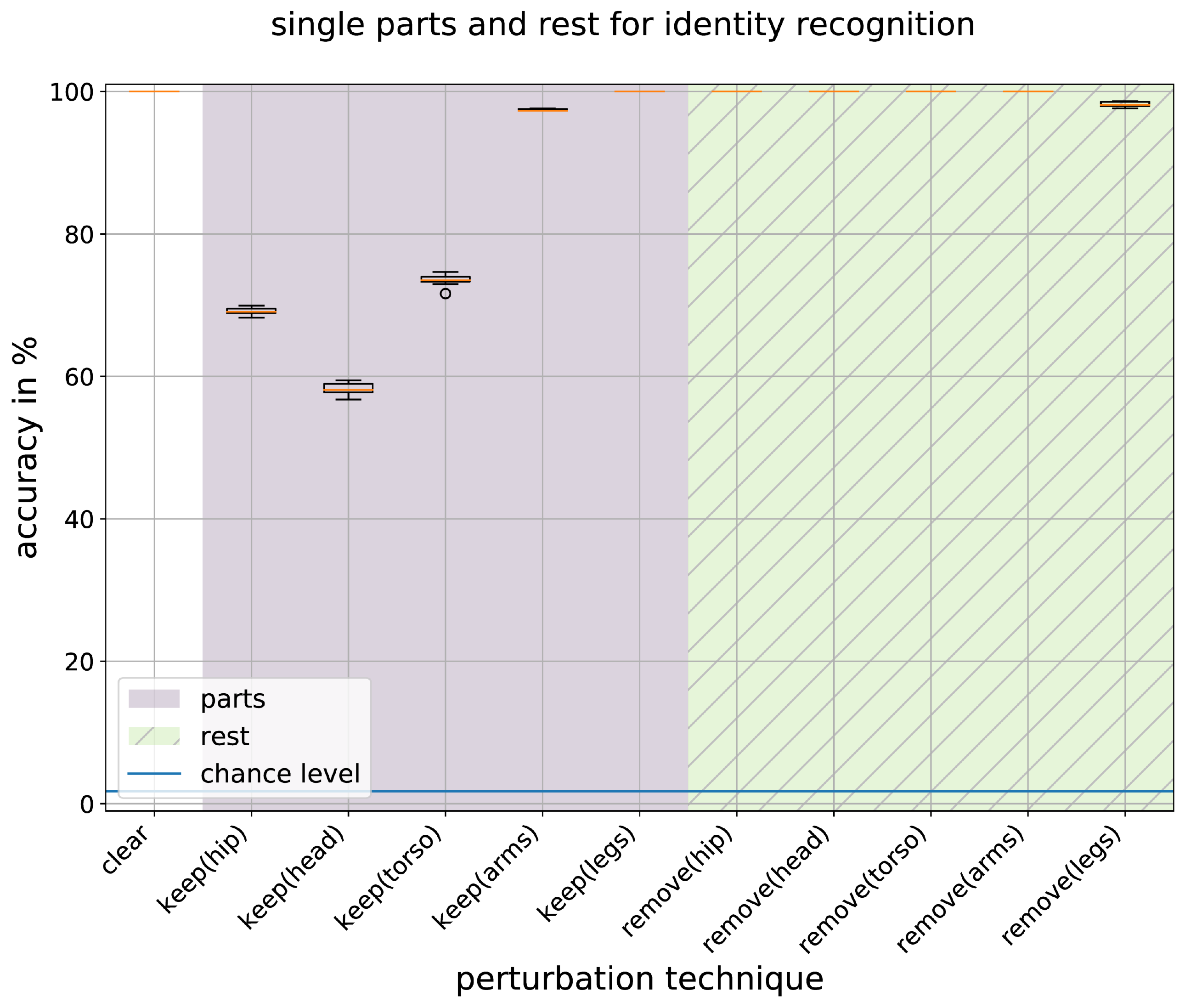}
\caption{Accuracy boxplots of results for individual body parts and all of the remaining body parts for identity recognition given in percent.}\label{img:parts_vs_all_identity}
\end{figure}

\begin{figure}[!h]
\centering
      \includegraphics[width=0.475\textwidth]{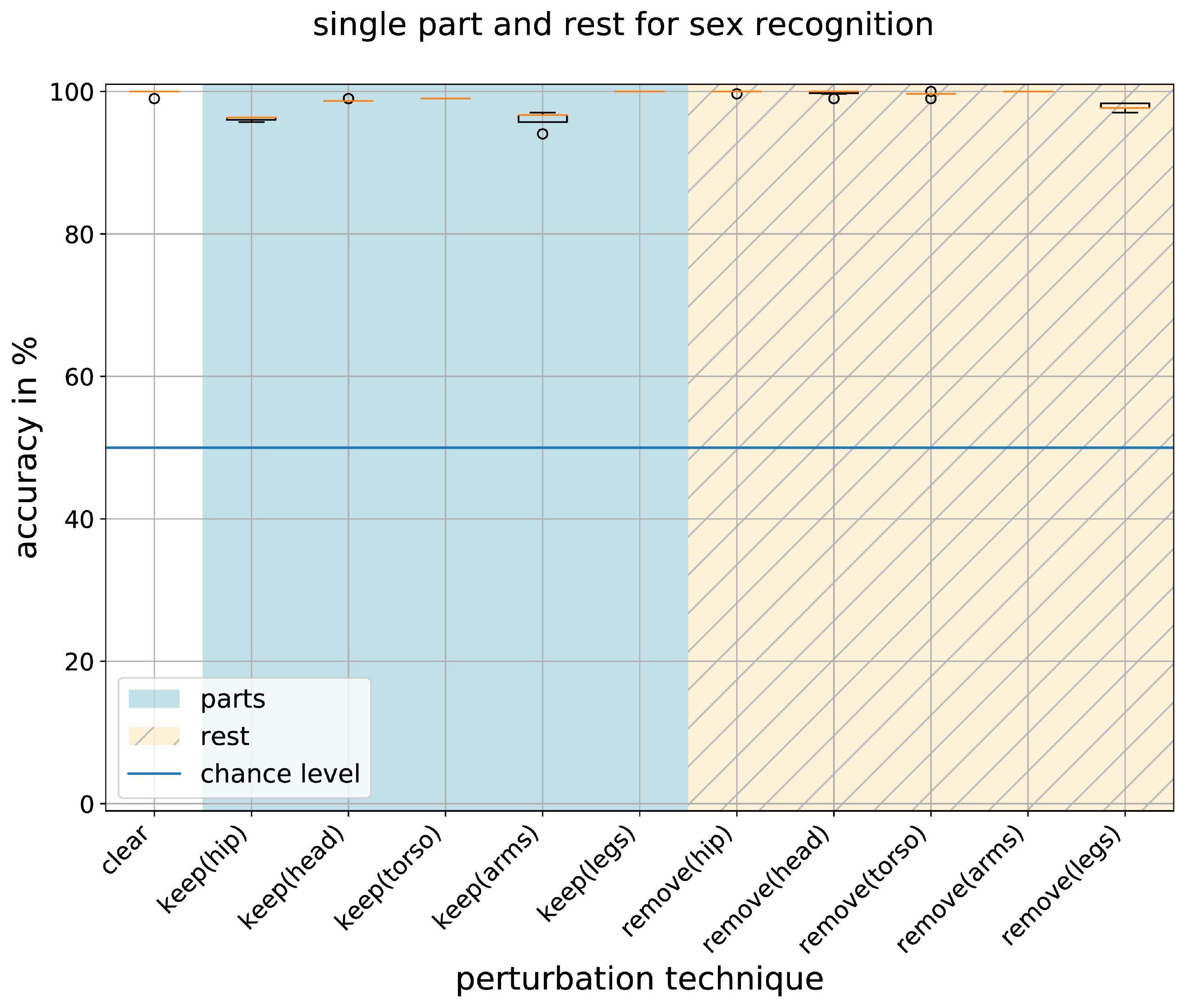}
\caption{Accuracy boxplots of results for individual body parts and all of the remaining body parts for sex recognition given in percent.}\label{img:parts_vs_all_gender}
\end{figure}

Next, we evaluate perturbations of individual isolated body parts in contrast to reduced whole body configurations (i.e., certain body parts were removed)
(see Fig.~\ref{img:parts_vs_all_identity} and Fig.~\ref{img:parts_vs_all_gender}). On the one hand, only the specified body part is used for the recognition (``keep''), while on the other hand, the whole body minus the specific body part (``remove'') is employed. Fig.~\ref{img:parts_vs_all_identity} shows that the removal of the legs slightly reduces the identity recognition accuracy to 97\%. At the same time, it is the only body part that achieves 100\% recognition accuracy alone. In contrast, keeping only the head as the standalone body part achieves the strongest prevention from identity recognition, reducing the accuracy to less than 60\%. Only slightly improved performance is achieved by the standalone body parts torso or hip.

Results for sex recognition with respect to perturbed body parts are displayed in Fig.~\ref{img:parts_vs_all_gender}. We find the same small reduction in accuracy for the removal of the legs as we saw for identity recognition, while it is again the only body part to achieve the full recognition accuracy as a standalone body part. However, for the other body parts, we find that their removal does not impact the sex recognition score. Additionally, our data shows only small effects on using only individual body parts in isolation. Comparing identity to sex recognition, head, hip, and torso alone fare much better for sex than for identity recognition.\textcolor{black}{ These results suggest that even the limited form information which is integrated over time into dynamic form information is sufficient to identify biological traits such as sex or even identity.
This finding is in line with human perception research. For example, }Kozlowski et al.~\cite{Kozlowski1977_gender} found that longer strides are perceived as more masculine. Center of moment contains sex information (see also Section \ref{sec:humanperception}; \cite{Kozlowski1977_gender,Montepare1988_age}).
\textcolor{black}{That is, as long as the stimuli contains information about certain body parts, sex and identity recognition is possible.}


\subsection{Dynamic vs. \textcolor{teal}{S}tatic \textcolor{teal}{F}eatures }

\begin{figure}[!h]
\centering
      \includegraphics[width=0.475\textwidth]{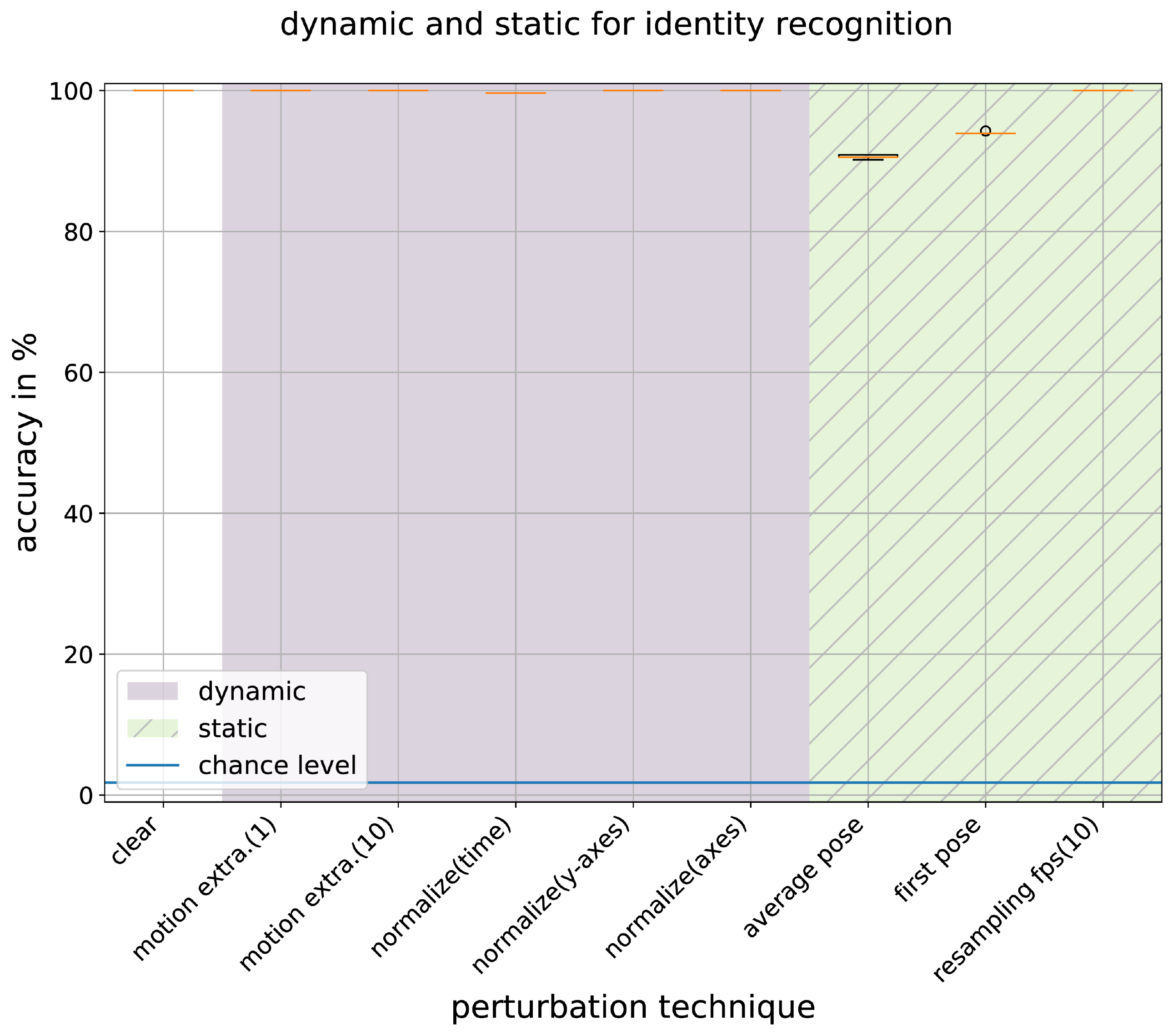}
\caption{Boxplots of accuracy results for dynamic and static features for identity recognition given in percent.}\label{img:dynamic_vs_static_identity}
\end{figure}

\begin{figure}[!h]
\centering
      \includegraphics[width=0.475\textwidth]{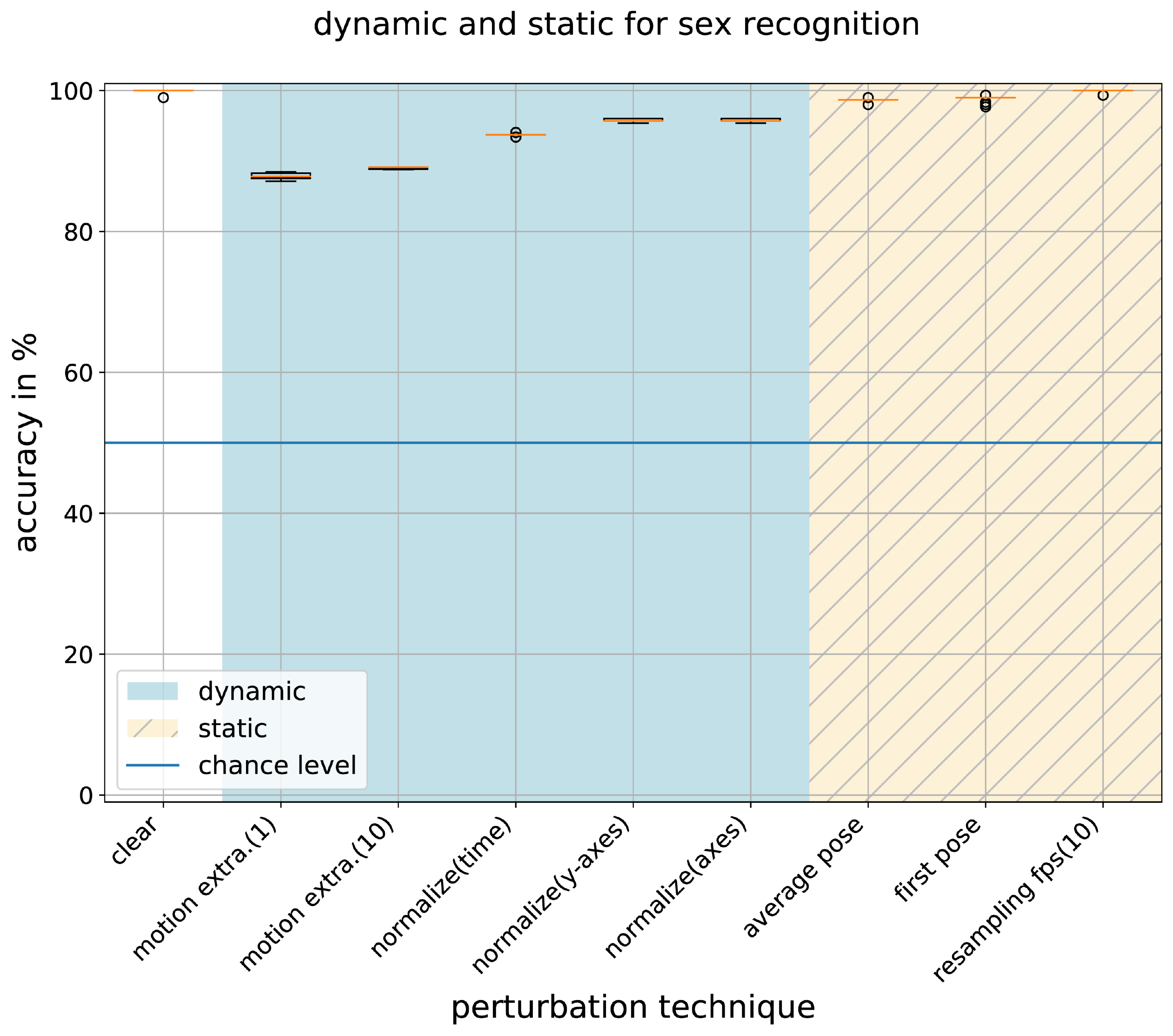}
\caption{Accuracy boxplots of results for dynamic and static features for sex recognition given in percent.}\label{img:dynamic_vs_static_gender}
\end{figure}

\textcolor{black}{Thirdly}, we investigate the effects of dynamic and static feature perturbation on recognition performance. In the case of identity recognition, depicted in Fig. \ref{img:dynamic_vs_static_identity}, we observe that only using the average pose or the first pose reduces the recognition accuracy slightly to 91\% and 94\% respectively, while other feature manipulations show no effect on identity recognition. For sex recognition (Fig.~\ref{img:dynamic_vs_static_gender}), our results show that while \textcolor{black}{static} features have close to no effect on accuracy, all \textcolor{black}{dynamic} features appear to do so. \textcolor{black}{So we can conclude that the static features are more important for the sex recognition than the dynamic ones.}
\newline

\subsection{\textcolor{black} {Combination of \textcolor{teal}{F}eatures}}

\textcolor{black}{
Here, we evaluate the combination of selected perturbation techniques from each category. Due to the further removal of data, we expect\textcolor{teal}{ed} to see larger reductions in the classification accuracy for both identity and sex recognition. We also expect\textcolor{teal}{ed} that with fewer data available the classification process becomes more unsteady and therefore the variance between the results will be larger. Further, the reduction of data can lead to a simplification of the data, which then is easier to classify.
}

\textcolor{black}{The combination of body parts head and leg\textcolor{teal}{s} with the static, dynamic, micro, and macro categories for sex recognition are shown in Fig.~\ref{img:parts_vs_all_gender}. Most of the legs combinations remain at 100\% accuracy. Only in combination with average pose and coarsening micro (100), a slight decrease in accuracy can be observed. When the legs are combined with coarsening macro (1000) we observe a large decrease in accuracy to close to 40\%, while both of these perturbations alone do not have an effect on the accuracy. The head (head alone achieves 60\% identity recognition) combinations are more of a mixed bag. While average pose and coarsening macro further reduce the accuracy\textcolor{teal}{;} resampling, coarsening micro, and remove variations do not have an additional effect on the accuracy. However, motion extraction, time normalization, and remove trajectory lead to an increase in the recognition accuracy. All three methods focus more on the smaller variations in the data, \textcolor{teal}{providing an} indication that the identification of individuals via their head motion is more dependent on the dynamic parts than the general movement.
}

\begin{figure}[!h]
\centering
      \includegraphics[width=0.475\textwidth]{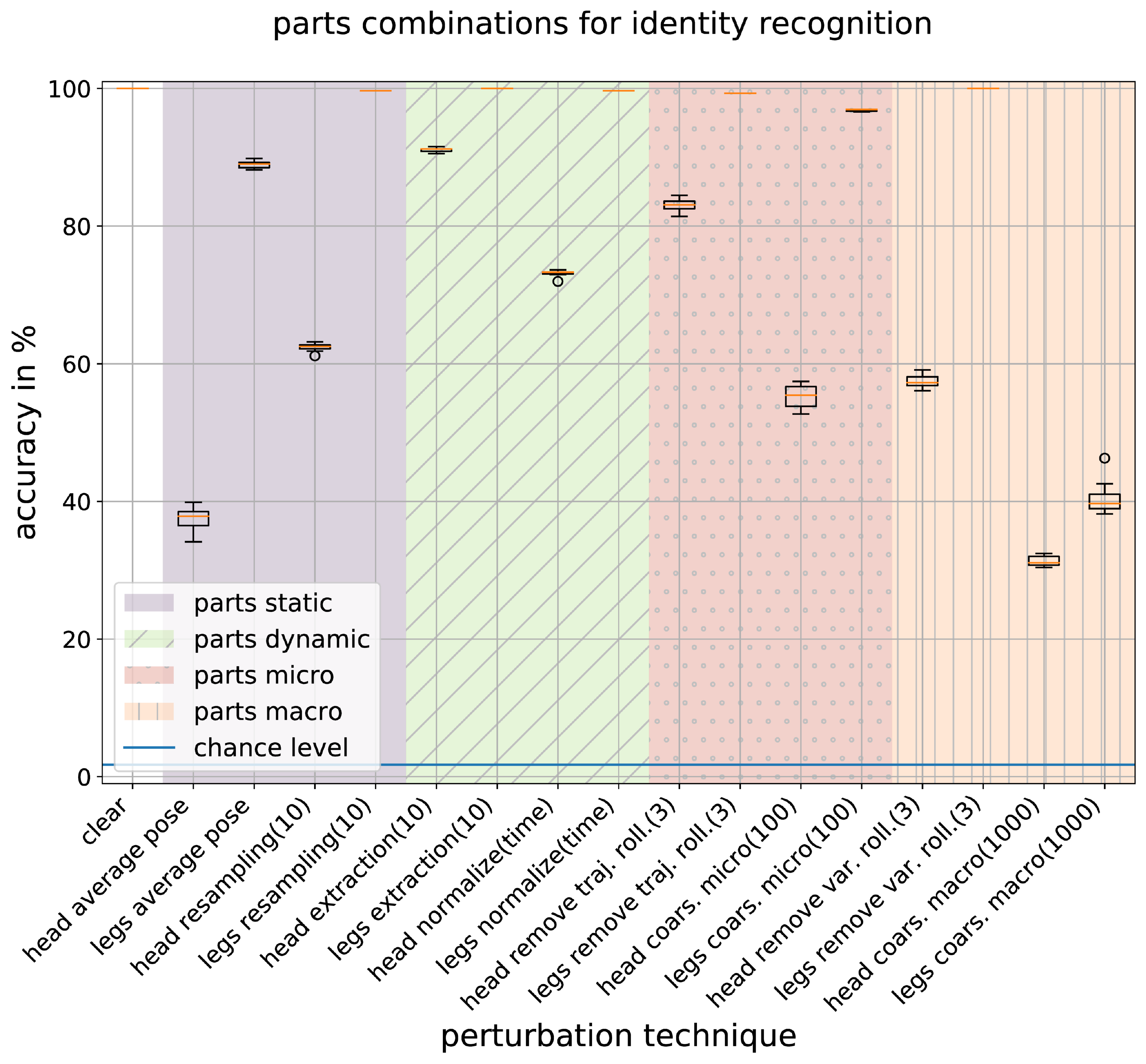}
\caption{Accuracy boxplots of results for legs and head in combination with the other categories for identity recognition given in percent.}\label{img:parts_combination_identity}
\end{figure}

\textcolor{black}{
\textcolor{teal}{Focusing on the combinations with head and legs for sex recognition, we find that while there is no effect on accuracy in combination with static features, the combination with dynamic features has deleterious effects on accuracy. Specifically, the combination of head and motion extraction results in a drop of accuracy to 75\%.}
For the micro combinations, we again find that the combinations with the head suffer the largest accuracy reduction. Here the combination with micro coarsening nearly reaches chance level, while the same combination with the legs stays above 90\%. When we compare this with the macro coarsening of the macro features\textcolor{teal}{,} we find that the legs drop to a lower accuracy than the head. This leads us to conclude that the sex recognition via the head data is much more dependent on the macro part of the head position, while the sex recognition via the legs depends more on the micro part of the positions.
}

\begin{figure}[!h]
\centering
      \includegraphics[width=0.475\textwidth]{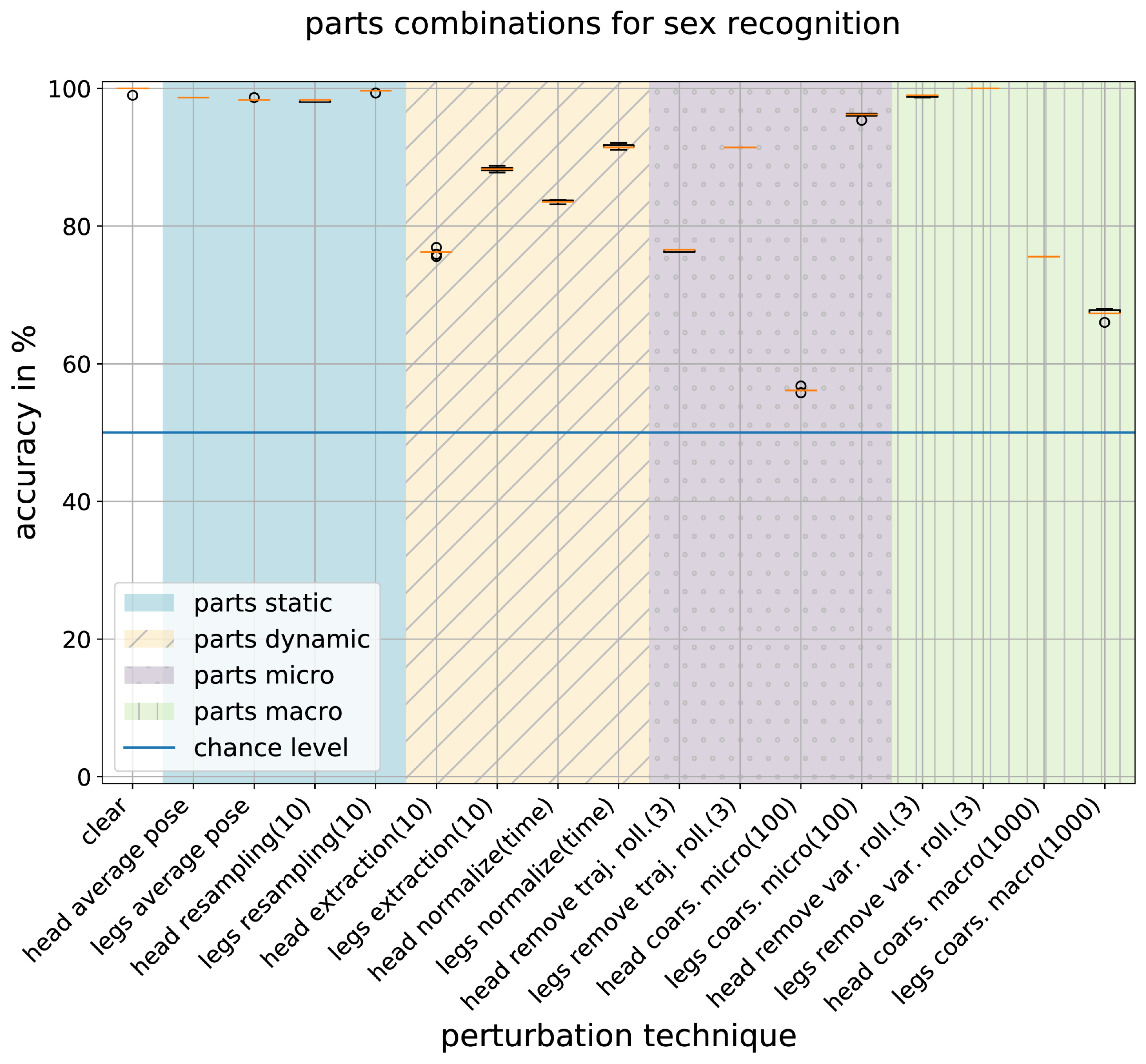}
\caption{Accuracy boxplots of results for legs and head in combination with the other categories for sex recognition given in percent.}\label{img:parts_combination_sex}
\end{figure}

\begin{figure}[!h]
\centering
      \includegraphics[width=0.475\textwidth]{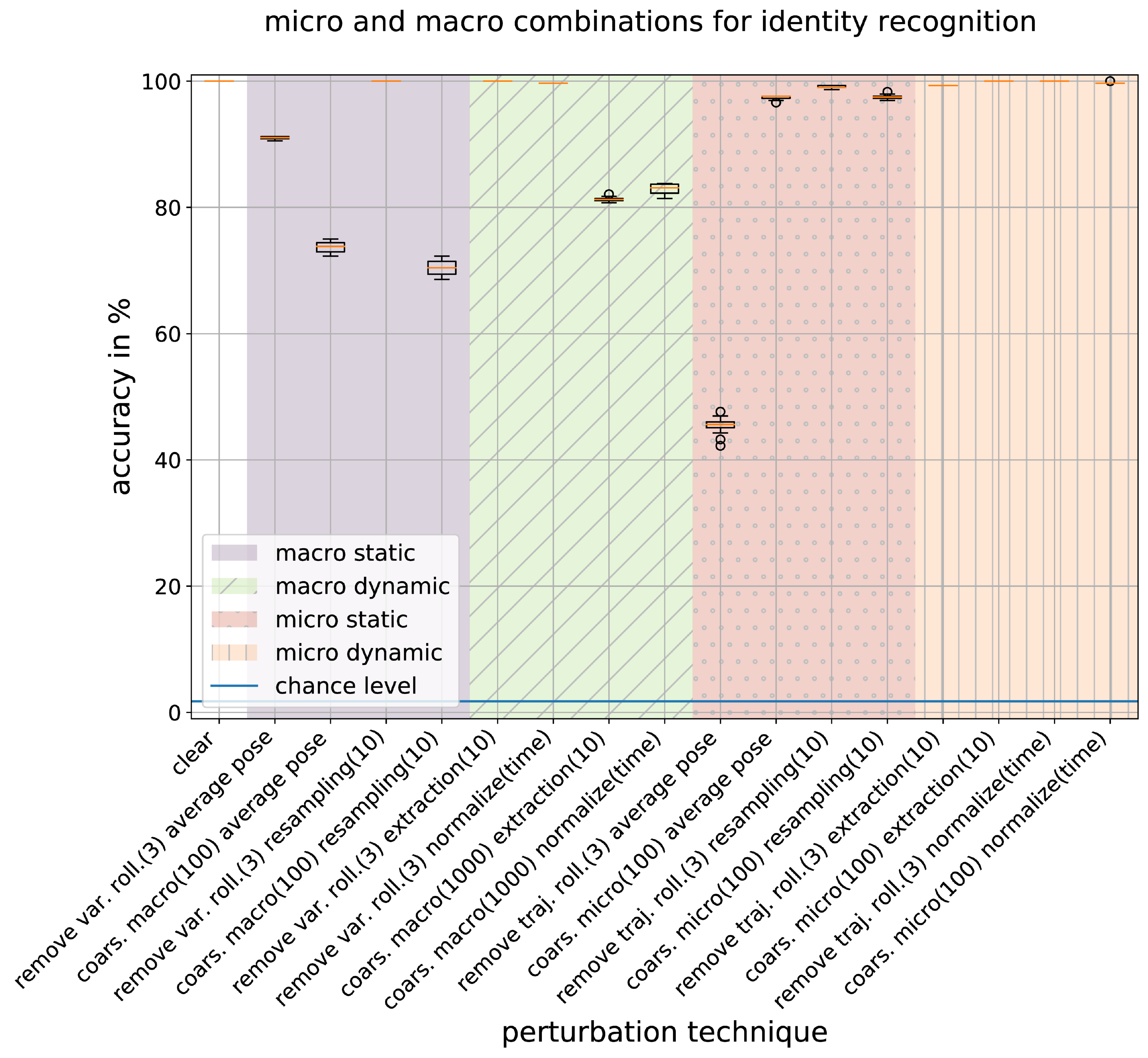}
\caption{Accuracy boxplots of results for macro, micro, dynamic, and static combinations for identity recognition given in percent.}\label{img:micro_macro_combination_identity}
\end{figure}

\textcolor{black}{
The results of macro, micro, dynamic, and static feature combinations for identity recognition are shown in Fig.~\ref{img:micro_macro_combination_identity}. The macro\textcolor{teal}{-}static combinations show accuracy decreases for the combinations that contain macro coarsening. We also see these decreases when we look at the combination of macro dynamic features in which the macro coarsening leads to a decrease in performance. The last combinations show an accuracy decrease in the removal of the trajectory plus average pose which drops the recognition accuracy to 45\%. Comparing the removal of variations and trajectory in combination with the average pose\textcolor{teal}{,} show that the general trajectory of the walker contains much identifiable information in their overall characteristic, while the small variations from the trajectories are only meaningful when their dynamic \textcolor{teal}{features are preserved}. 
}

\begin{figure}[!h]
\centering
      \includegraphics[width=0.475\textwidth]{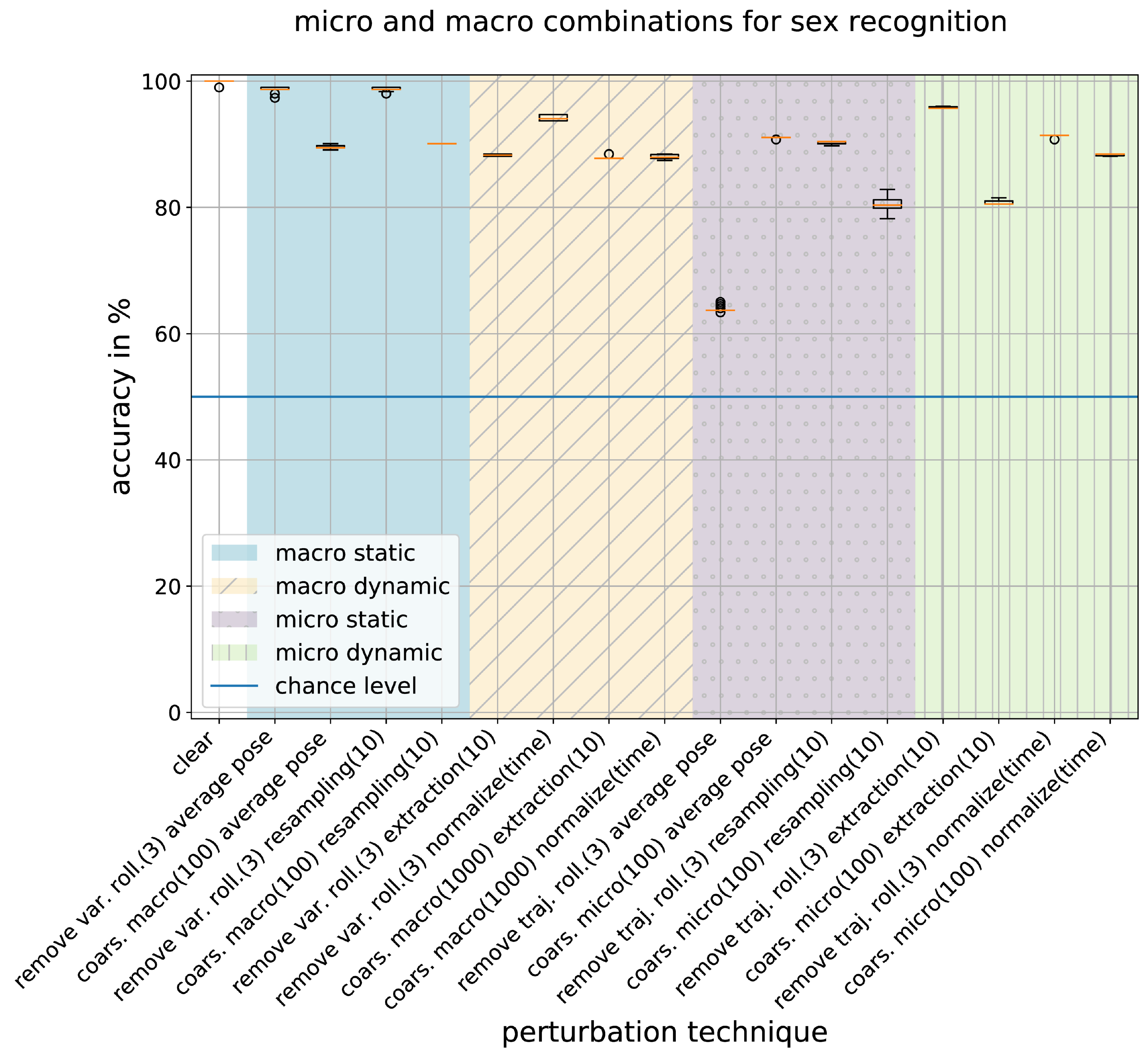}
\caption{Accuracy boxplots of results for macro, micro, dynamic, and static combinations for sex recognition given in percent.}\label{img:micro_macro_combinations_gender}
\end{figure}

\textcolor{black} {
Lastly, we look at the same feature combinations as before but this time for sex recognition (see Fig.~\ref{img:micro_macro_combinations_gender}). \textcolor{teal}{In the case of t}he combination of macro and static features\textcolor{teal}{,} the removal of the variations does not lead to a drop in accuracy, while both combinations with coarsening macro drop to the same accuracy level of about 90\%. \textcolor{teal}{This suggests that obfuscations in combination with macro features have a bigger impact on the accuracy in comparison to combinations with static features. All macro-dynamic combinations result in a decrease of performance to about 90\%.}
Furthermore, removing the variations plus macro coarsening increases the performance slightly when compared to just performing the same macro coarsening alone \textcolor{teal}{(see Fig.~\ref{img:micro_vs_macro_gender})}. In the micro\textcolor{teal}{-}static combinations, we find the removal of the trajectory average pose combination to create a \textcolor{teal}{large drop in} accuracy.
}

\subsection{\textcolor{black} {Utility}}

\begin{figure}[!h]
\centering
      \includegraphics[width=0.475\textwidth]{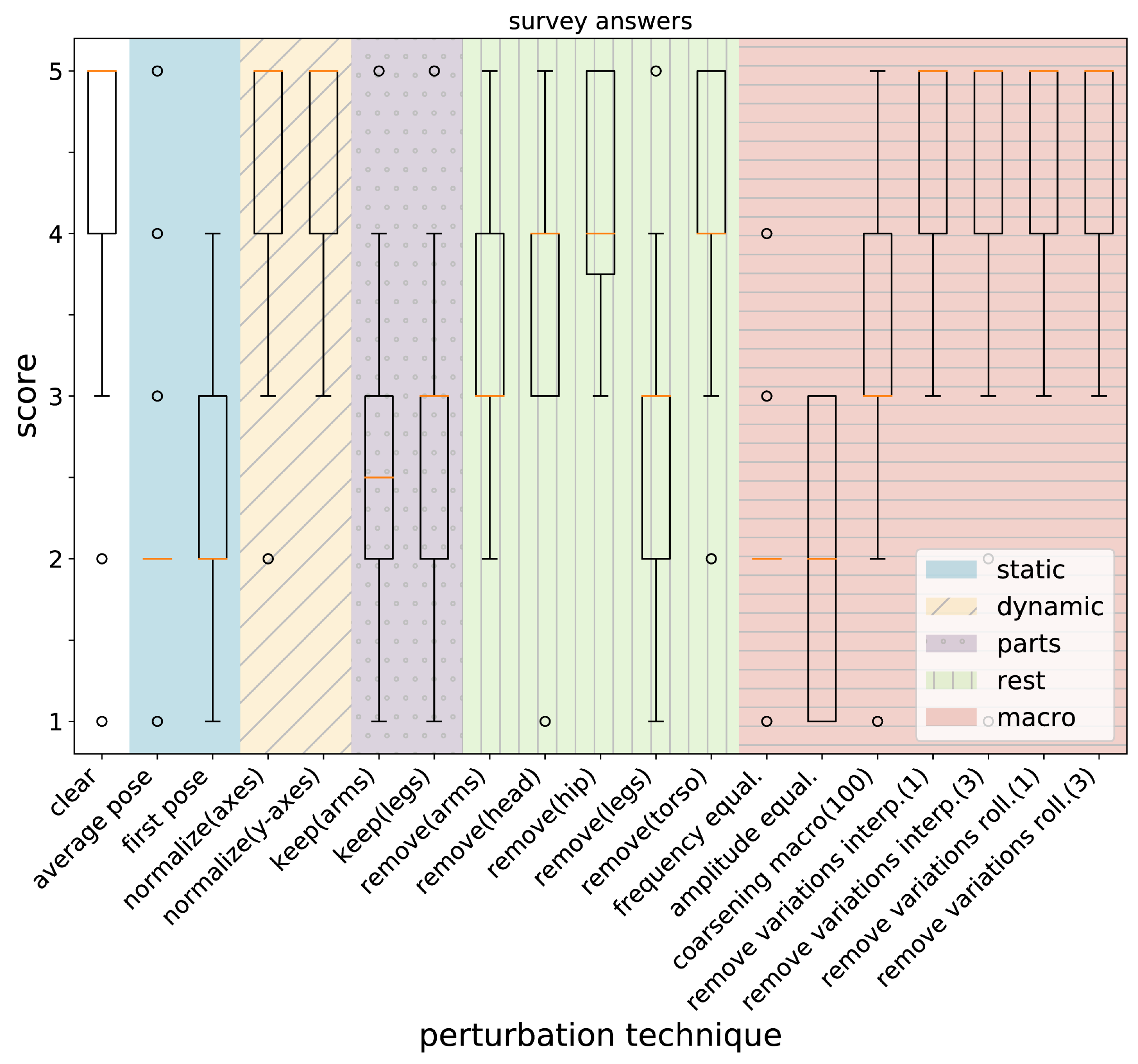}
\caption{Boxplots of the naturalness rating scores for the perturbation techniques that retained utility.} 
\label{img:survey_responses}
\end{figure}

Finally, we report the results of the naturalness evaluation for all perturbation techniques that have a median rating score which is greater than 1 (all techniques that retain some utility; \textcolor{teal}{on the 1-5 scale described in Sect.~\ref{sec:meth_utility}) and are shown} in Fig.~\ref{img:survey_responses}. Perturbations that resulted in a median score below that, were assumed to retain no utility and are therefore not plotted. First, we note that \textcolor{teal}{none of the micro feature perturbations retained any gait naturalness.} 
In the static category only average and first pose \textcolor{teal}{managed to appear minimally natural, with median naturalness scores of 2. The exclusion of body parts of the walkers had deleterious effects on the perceived naturalness, while still maintaining some level of naturalness depending on the specific removed body part. Interestingly, keeping only the arms or legs of the walker was rated as still somewhat natural, whereas all other individual body parts in isolation were rated as non-natural.}
The normalization of all axes and \textcolor{teal}{the normalization of} the y-axes \textcolor{teal}{achieve} the same level (median of 5) of naturalness as the clear data. The only other techniques that achieve the same naturalness \textcolor{teal}{ratings} are the remove variation techniques. In general, these results are within our expectations\textcolor{teal}{,} as perturbing the data should either \textcolor{teal}{maintain} the same level of naturalness or decrease it. The fact that most of the macro features retained the naturalness of the walker is also unsurprising\textcolor{teal}{,} as they preserve the majority of the gait variations while the small variations we kept in the micro features are not perceived as natural anymore. We did not evaluate the naturalness of the combinations, however, we assume that a combination will at most reach the minimum naturalness \textcolor{teal}{rating score} of its two used perturbation techniques.

\section{Discussion}\label{sec:discussion}
\textcolor{black}{Using ML for gait recognition based on motion capture data, we investigated the importance of features based on findings in psychology for identity and sex recognition. The findings reported here, suggest that all of the features reported by psychology are transferable to ML approaches in identification performance based on walking motion. The identification procedure is robust as even when large parts of the data are removed the identification rates are high, only when multiple features are removed from the data a significant impact on the accuracy can be observed.
Consistent with previous studies in psychology and neuroscience \cite{Peng2020,YovelOToole2016,Troje2005}, we found that dynamic and static features contain much identifiable information, hinting at strong temporal and physiological dependencies in the data.}
\textcolor{black}{
We anticipate that for the development of suitable anonymization techniques for gait data the dependencies between the features have to be accounted for, as otherwise, the reconstruction of the clear data is likely possible. 
For example, noise that is applied to the marker positions could be removed by smoothing the trajectories, or missing markers can be reconstructed from the position of the remaining ones.
Wang et al.~\cite{wang21why} have convincingly demonstrated this, showing how adding noise does not effectively perturb correlated data.
}
\textcolor{black}{
Interestingly, the removal of body parts and the subsequent performance accuracy alone indicates a high redundancy in the data, and as such focusing on a single feature for anonymization is unlikely to achieve a meaningful anonymization effect. This effect, albeit in a much weaker form,  has previously been shown in human person and biological motion perception studies: The elimination of some local information, for example by removing \acrshort{pld} dots corresponding to body parts, does not affect the recognition as long as a certain degree of global form revealing dynamic posture changes is preserved \cite{beintema2002perception,lange2006visual}.
}


\textcolor{black}{
Both, the overall trajectories of the gait as well as small variations in the data, allow for recognition of individuals. Thus, making it necessary to adjust the overall gait trajectory for anonymization purposes.
The overall pattern of results here provides converging evidence for the need to consider gait motion capture a strong personal identifiable trait, even when recorded at low resolution or low frame rate. Many features, as investigated here — macro, micro, dynamic and static features as well as individual body parts — contain strong identifiable information about both, the identity and the sex of a human walker. With our simple ML-based feature perturbation approach we found that coarsening the marker positions precision, with the respective recognition performances of 45\% and 2\% for sex and identity exhibited the strongest reduction of classification accuracy while removing dynamic \& static features generally only reduced recognition slightly. However, our utility evaluation of the features shows that the perceived naturalness of the perturbed data is diminished when the general motion or body structure of the walkers is removed.
Thus we see a strong indication that in order to develop strong anonymization for gait data, while keeping its utility intact, a holistic approach is required. Such an approach should take the dependencies in the data and the requirement for natural-looking results into account, for example by generating synthetic gait trajectories.
}


\subsection{Limitations and \textcolor{teal}{F}uture \textcolor{teal}{W}ork}\label{sec:future work}

The present work is based on a data set of 57 young adult individuals and as such it might be possible to achieve superior anonymization results for larger sample sizes. However, as we have shown gait data does contain a large amount of identifiable information, so larger effects from bigger samples are unlikely. The present work presents results on one sole gait cycle per \textcolor{black}{sample}, future work should include multiple {sequential} gait cycles or gait data from \textcolor{black}{multiple sessions}. Furthermore, all individuals were from a similar age cohort, including different lifespan age brackets or longitudinal data might lead to more meaningful and representative results. However, we believe that having a cohort of very similar individuals also strengthens the recognition results, as it becomes more difficult to tell the individuals apart. It is possible that with the improvements of machine learning approaches\textcolor{teal}{,} better classification results can be achieved on our perturbed data\textcolor{teal}{. A}s such our approach only gives a lower bound how much identifiable information remains in the perturbed data. \textcolor{black}{This fact is also shown by some of the combinations of perturbation techniques where the combinations achieved higher recognition accuracy than the individual techniques alone.} 

\textcolor{teal}{With regards to the user study we would like to point out that our definition of utility only takes into account how natural other people perceived the anonymized PLD gait sequences shown to them. We did not investigate if the original walker themself would find their perturbed gait to be natural. We did so because we assumed that the device used in our system-and-threats-model is trusted by the user and therefore would display the real gait (pre-transfer to the service provider described in Sect. \ref{sec:system_model}; labeled "clear" in the present work) to the user as it is recorded locally in real-time, instead of an anonymized version of the user's gait. Furthermore, we based our present investigation on an existing open-source dataset and therefore have no access in an ethical and legal way to the original walkers due to inter alia data protection and privacy reasons. Future studies that obtain their own motion capture recordings could include an evaluation of utility by asking the recorded walkers themselves to evaluate their perturbed gait or other movements recorded with motion capture. 
}

For \textcolor{teal}{additional} future work we propose to conduct the same set of experiments with human observers 
to directly compare human and machine gait recognition, in order to gain insight into how both differ \textcolor{teal}{in regards to identifying individuals and their sex}. Although, the human ability to process biological motion such as gait-based person perception and recognition is susceptible to viewer-specific influences such as age \cite{billino2019motion}, social factors (e.g., interpersonal context, stereotypes) \cite{bolling2013social,Johnson2011}, neurodevelopmental disorders (e.g., autism, schizophrenia) \cite{bolling2013social}, and other potential experimental
, concomitant
, and individual 
factors \cite{foster2019decoding,herrington2011biological,vanderZwan2009gender,connor2018biometric,Johnson2011,loffing2016tennishand}. Thus, utilizing machine gait recognition provides a more objective evaluation method for different anonymization techniques.


\section{Conclusion}\label{sec:conclusion}

In this paper, we address the question of how much specific features of human gait contribute to the ability to discern the identity or sex of different human individuals in gait data.
Here, we found that overall identification performance was indeed very robust. Removing large parts of the data, either by omitting body parts or reducing spatial and temporal resolution, did have little effect on the recognition performance.

One possible interpretation of the findings is that gait is idiosyncratic and very redundant. Moreover, gait can be considered an individual trait that shows little variability over time and even lifespan. Studies reported that major adult gait emerges already at the age of five years, although age-related effects such as slower gait or shorter steps as well as age-related body proportion changes have been found as well \cite{Montepare1988_age}.

Our results suggest that gait will be very hard to anonymize effectively.
This entails that anonymization cannot be achieved with simple means, but will require intricate approaches that take the inter \textcolor{teal}{-}dependency of the connected body, as well as the overall generating process of the walking human into consideration. \textcolor{teal}{Utility can only be retained when the macro structure of the walker and its dynamic are largely kept intact.}

\begin{acks}
\label{sec:Ackn}
This work has been supported by the German Research Foundation (DFG, Deutsche Forschungsgemeinschaft) as part of Germany's Excellence Strategy -- EXC 2050/1 -- Project ID 390696704 -- Cluster of Excellence ``Centre for Tactile Internet with Human-in-the-Loop" (CeTI) of Technische Universit{\"a}t Dresden, and by funding of the Helmholtz Association through the KASTEL Security Research Labs (HGF Topic 46.23).
\end{acks}
\printbibliography
\end{document}